\documentclass[sn-mathphys-num]{sn-jnl}% Math and Physical Sciences Numbered Reference Style 
\usepackage{graphicx}%
\usepackage{multirow}%
\usepackage{amsmath,amssymb,amsfonts}%
\usepackage{amsthm}%
\usepackage{mathrsfs}%
\usepackage[title]{appendix}%
\usepackage{xcolor}%
\usepackage{textcomp}%
\usepackage{manyfoot}%
\usepackage{booktabs}%
\usepackage{algorithm}%
\usepackage{algorithmicx}%
\usepackage{algpseudocode}%
\usepackage{listings}%
\usepackage{lscape}
\usepackage{rotating}
\usepackage{tabularx}
\usepackage{fancyhdr}
\newif\ifcomment
\commenttrue

\newcommand{\cmicrons}{$\mu\mathrm{m}^3$}
\newcommand{\micron}{$\mu\mathrm{m}$}
\newcommand{\Rfact}{\mathcal{R}}
\theoremstyle{thmstyleone}%
\theoremstyle{thmstyletwo}%

\theoremstyle{thmstylethree}%

\fancypagestyle{FirstPage}{
\lfoot{\copyright2024. All rights reserved.} 
}

\begin{document}
%\begin{bibunit}

\title[Article Title]{A 25-micron single photon sensitive kinetic inductance detector}


\author*[1]{\fnm{Peter K.} \sur{Day}}\email{Peter.K.Day@jpl.nasa.com}

\author[2]{\fnm{Nicholas F.} \sur{Cothard}}%\email{iiauthor@gmail.com}

\author[3]{\fnm{Christopher} \sur{Albert}}%\email{iiiauthor@gmail.com}

\author[3]{\fnm{Logan} \sur{Foote}}%\email{iiiauthor@gmail.com}

\author[3]{\fnm{Elijah} \sur{Kane}}%\email{iiiauthor@gmail.com}

\author[1]{\fnm{Byeong H.} \sur{Eom}}%\email{iiiauthor@gmail.com}

\author[1]{\fnm{Ritoban} \sur{Basu Thakur}}%\%email{iiiauthor@gmail.com}

\author[1]{\fnm{Reinier M.J.} \sur{Janssen}}%\email{iiiauthor@gmail.com}

\author[1]{\fnm{Andrew} \sur{Beyer}}%\email{iiiauthor@gmail.com}

\author[1]{\fnm{Pierre} \sur{Echternach}}%\email{iiiauthor@gmail.com}

\author[1]{\fnm{Sven} \sur{van Berkel}}%\email{iiiauthor@gmail.com}

\author[3]{\fnm{Steven} \sur{Hailey-Dunsheath}}%\email{iiiauthor@gmail.com}

\author[2]{\fnm{Thomas R.} \sur{Stevenson}}%

\author[4]{\fnm{Shahab} \sur{Dabironezare}}

\author[4]{\fnm{Jochem J.A.} \sur{Baselmans}}%\%email{iiiiauthor@gmail.com}

\author[2]{\fnm{Jason} \sur{Glenn}}%\email{iiiauthor@gmail.com}

\author[1]{\fnm{C. Matt} \sur{Bradford}}%\email{iiiauthor@gmail.com}

\author[1]{\fnm{Henry G.} \sur{Leduc}}%\email{iiiauthor@gmail.com}

\affil*[1]{\orgdiv{Jet Propulsion Laboratory}, 
\orgname{California Institute of Technology}, \orgaddress{\street{4800 Oak Grove Dr.}, \city{Pasadena}, \postcode{91107}, \state{CA}, \country{USA}}}

\affil[2]{%\orgdiv{}, 
\orgname{NASA Goddard Space Flight Center}, \orgaddress{\street{8800 Greenbelt Rd}, \city{Greenbelt}, \postcode{20771}, \state{MD}, \country{USA}}}

\affil[3]{%\orgdiv{}, 
\orgname{California Institute of Technology}, \orgaddress{\street{1200 California Blvd.}, \city{Pasadena}, \postcode{91125}, \state{CA}, \country{USA}}}

\affil[4]{%\orgdiv{}, 
\orgname{SRON Netherlands Institute for Space Research}, \orgaddress{\street{Niels Bohrweg 4}, \city{Leiden}, \postcode{2333 CA}, 
\country{The Netherlands}}}

\abstract{We report measurements characterizing the performance of a kinetic inductance detector array designed for a wavelength of 25 microns and very low optical background level suitable for applications such as a far-infrared instrument on a cryogenically cooled space telescope.  In a pulse counting mode of operation at low optical flux, the detectors can resolve individual 25-micron photons.  In an integrating mode, the detectors remain photon noise limited over more than six orders of magnitude in absorbed power from 70~zW to 200~fW, with a limiting NEP of $4.6\times10^{-20}~\rm W\,Hz^{-1}$ at 1~Hz. In addition, the detectors are highly stable with flat power spectra under optical load down to 1~mHz.  Operational parameters of the detector are determined including the efficiency of conversion of the incident optical power into quasiparticles in the aluminum absorbing element and the quasiparticle self-recombination constant.}


\thispagestyle{FirstPage}


\maketitle

\section{Introduction}\label{sec1}

\textcolor{black}{The thermal infrared through millimeter-wave band ($\sim 5-2000 \mu$m) has great promise for breakthrough measurements in fields ranging from exoplanet spectroscopy, planet formation, galaxy evolution, and cosmology.  Relative to the optical bands, the thermal IR ($\sim5-25 \mu$m) offers relaxed contrast between planets and their host stars and carries spectral features of many biogenic gases.  This band is being developed for transit spectroscopy with JWST \cite{Dyrek_2024,Bouwman_2023} and future concepts are targeting direct exoplanet spectroscopy via interferometric nulling from space \cite{quanz_21}.  In the far-IR ($\sim$25--300$\mu$m), massive gains are possible for general astrophysics, including planet formation and galaxy evolution, with a cryogenic space-borne telescope \cite{Bradford_21,Glenn_2024}.   In the submillimeter and millimeter ($\sim$300--2000$\mu$m), wide-field imaging spectroscopy creating tomographic maps of the Universe provides unique constraints at the intersection of cosmology \cite{Karkare_22} and galaxy evolution in the first billion years after the Big Bang \cite{Gong_2012,Sun_2021}. } 

Realizing these opportunities, however, requires high-performance detector arrays which are not available from commercial providers.  Exoplanet transit work requires  excellent stability ($\sim$10~ppm or better) over many hour timescales \cite{Sakon_2021} to accurately difference the planet from the star. Far-IR astrophysics and millimeter-wave cosmology require exquisite per-pixel sensitivity (noise equivalent power NEP $\lesssim 10^{-19}\,\rm W\,Hz^{-1/2}$) to match the natural backgrounds of a cryogenic space-borne platform, combined with high dynamic range ($\sim10^5$) to accommodate the diversity of sources.  All applications require many-kilopixel array formats, necessitating efficient multiplexing.

Superconducting pair breaking detectors, including kinetic inductance detectors (KIDs) \cite{Day_2003,Baselmans_17}, superconducting tunnel junction (STJ) detectors \cite{peacock1997superconducting} and quantum capacitance detectors (QCDs) \cite{Echternach_18}, are a powerful emerging approach for this next generation of observatories and instruments. These devices operate by sensing changes in the Cooper pair density through the quasiparticle population produced by absorbed photons.  Owing to the small value of superconducting gap energy, typically hundreds of $\mu$eV, these detectors can operate through the far-IR and much of the millimeter-wave bands; aluminum, for example, has a pair-breaking energy threshold corresponding to 90~GHz.  In the far-infrared through millimeter, the devices typically work as total power detectors: fluctuations in the absorbed power create corresponding fluctuations in the quasiparticle density.  Recent efforts have yielded both KIDs and QCDs with noise equivalent power better than 10$^{-19}\,\rm W\,Hz^{-1/2}$ \cite{Baselmans_22,Echternach_18,Hailey_23,foote_high-sensitivity_2024}, meeting the requirements for space-borne moderate-resolution spectroscopy in the 200--300~\micron\ regime.  At shorter near-IR through visible wavelengths, each absorbed photon breaks enough Cooper pairs to produce a measurable pulse response, so the detector can operate in a photon-counting mode where the detected pulse amplitudes indicate the photon energies, and the arrival times can be determined \cite{mazin_2019}.  KIDs in particular have emerged as a versatile detectors, given successes in ground-based and balloon-borne instrumentation in both the far-IR to millimeter \cite{Sayers_2014,Bing_2023_NIKA2_cosmology,Calvo_2016_NIKA2,Ritacco_2017_NIKAPol,Endo_2019Deshima,Presta_2020_Olimpo,Lowe_2020_BlastTNG}, and the optical \cite{Meeker_2018_Darkness,Mazin_2013_Archons}, as well as advances in readout electronics \cite{Bourrion_2016_NIKAreadout,vanRantwijk_2016_SRONreadout,Gordon_2016_ASUreadout,sinclair_ccat-prime_2022}.

\textcolor{black}{Here we describe a KID demonstration at an intermediate, mid-IR wavelength of 25~\micron, which is of prime interest for future spaceborne astrophysics experiments.} It operates in both integrating and photon-counting modes depending on the optical flux.  At low arrival rates, individual 25-micron photons are detected and can be distinguished from other events, such as cosmic rays or background radiation, based on the pulse amplitudes.  %\textcolor{black}{The photon counting allows for excellent calibration of device response at very low power levels.}  
At higher arrival rates, the detector acts as a shot-noise-limited integrating detector over a dynamic range exceeding six orders of magnitude in optical power.  In that mode, the detector is stable on the time scale of hours, allowing for deep integration to measure small fractional changes in optical flux.  \textcolor{black}{Section~\ref{sec:device} provides a brief description of the detector, Section~\ref{sec:results} presents the principal results, and Section~\ref{sec11} outlines the methods used. Sections~\ref{sec:discussion} and \ref{sec:conclusion} provide context and summarize.}

%\ifcomment
%\textcolor{red}{Include some motivation for detector stability, both as could be applied to a mission like PRIMA and also possibly an exoplanet transit type measurement. }
%\fi

\textcolor{black}{\section{Device}\label{sec:device}}

KIDs operate by sensing the effect of optically excited quasiparticles on the complex conductivity of the inductive element of a superconducting microresonator.  At far-IR wavelengths, efficient optical coupling has been achieved using either wire grid absorbers \cite[e.g.]{Hailey_18} or various antenna structures \cite[e.g.]{Baselmans_17}.  Both of these techniques are compatible with low inductor volumes needed to enhance the detector sensitivity, particularly when used in concert with a microlens array or feedhorn array. 
%In the case of absorber coupling,  the absorber area can be decreased through the use of a microlens array or feedhorn to focus the incoming light. 
Here we present a KID design for 25-microns, around the short wavelength end of the far-infrared band, where antenna coupling becomes inefficient due to conductor loss \textcolor{black}{and feedhorn manufacture is not practical due to surface roughness requirements.  These considerations drive us to a microlens / absorber coupled approach.    While many such devices for submillimeter to optical frequencies have used high-normal-state-resistivity superconductors which allow straightforward resistive absorption, our device builds on our earlier work showing that an absorber incorporating resonant structures \cite{cothard_AlPPCKIDs,Nie_22} can have high efficiency in the far-IR even when using low-resistivity aluminum film, now considered the material of choice for most KID applications.}
%At wavelengths down to ???-microns and at optical frequencies absorber-coupled KIDs based on high-normal-state-resistivity superconductors have been used.  In earlier work, we showed that an absorber incorporating resonant structures\cite{cothard_AlPPCKIDs} made from a low resistivity aluminum film could be efficient at 25-microns and results in a very low inductor volume.  
Our array builds on that presented in \cite{cothard_AlPPCKIDs}, but incorporates both further reduced volume and improved quasiparticle lifetime.  These factors combine to create a detector that is photon shot-noise-limited under the aW level optical loading predicted for space-borne far-infrared spectroscopy.  %By counting individual photons we are able to precisely calibrate the optical power at this low level.

\begin{figure}[h]
\centering
\includegraphics[width=1.00\textwidth]{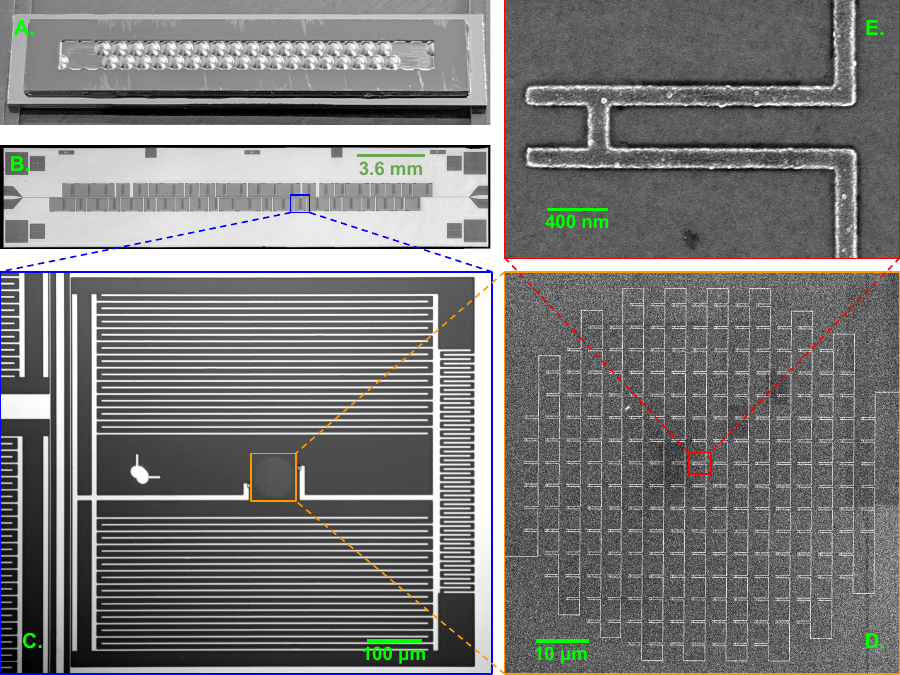}
\caption{Photographs and scanning electron microscope (SEM) image of the KID array and components. A: Photograph of anti-reflection coated microlens array. B: stitched microscope photographs of the 44-element kinetic inductance detector array. C: A single KID comprised of a pair of niobium IDCs (upper and lower) surrounding the 70 micron diameter aluminum absorber (center). The detector is capacitively coupled to a niobium feedline (left) and groundplane (right). D: SEM image of the meandered aluminum resonant absorber. E: SEM image of the periodically patterned ``hairpin'' geometry resonant structure, showing 100 nm linewidth aluminum trace width. The resonant absorber is discussed further in Appendix \ref{sec:res}.}
\label{fig:pixel}
\end{figure}

The detector array consists of 44 KIDs arranged on a hexagonal lattice with a pitch of 900 microns.  The layout of a pixel is shown in Fig.~\ref{fig:pixel}.  It consists of a 70-micron diameter aluminum meandered inductive section in parallel with a niobium interdigitated capacitor (IDC) forming a resonator.  The resonance frequencies were distributed log-uniformly in the range of 570 to 1120~MHz.  Coupling quality factors $Q_c$ were distributed in the range of $2\times 10^4$ to $10^5$ and the unilluminated internal quality factors were $Q_i \approx 10^6$. Details on the array uniformity are described in Appendix~\ref{sec:uniformity}. The array is fabricated on a high resistivity silicon wafer with a similar process to that described in \cite{Hailey_23}, except that the aluminum film thickness is 40~nm and the line width of the inductor is reduced to 100~nm in order to minimize its volume.  The meander is similar to the one described in \cite{cothard_AlPPCKIDs} except for narrower line width, and is made up of the repeated unit cell shown in panel (E) of the figure.  The pattern, which acts as a frequency selective surface, is designed to resonantly absorb incident light in a band around 25 microns in both polarizations.  The optically generated quasiparticles are confined to the aluminum because of the higher niobium gap parameter.  We estimate an aluminum volume of 11~\cmicrons, taking into account that as fabricated, the lines were slightly wider than designed.  Light is focused on the aluminum absorbers using an anti-reflection coated silicon lenslet array that is glued to the back side of the detector wafer using a submicron thickness epoxy layer, as described in \cite{cothard_microlenses}.  Eight of the pixels were left without lenses.

\section{Results}\label{sec:results}

The measurements were conducted using a standard homodyne circuit to probe the complex transmission coefficient $S_{21}$ of a particular resonator.  The microwave probe power was set to be approximately 10~fW, roughly 3~dB less than the bifurcation power of the resonators, which is the power at which the resonance curves becomes hysteretic \cite{swenson2013operation} due to the nonlinear response of the kinetic inductance.  The array was kept at a temperature of 150~mK for most of the measurements. Characterization of the detectors at other bath temperatures is described in Appendix~\ref{sec:basetresp}.

Changes in $S{21}$ are converted to changes in the fractional frequency shift $\delta x = \delta f_r / f_r(0)$, where $f_r(0)$ is the average resonance frequency without optical illumination and $\delta f_r$ is the resonance frequency perturbation.  The detectors are exposed to optical power from a cryogenic black body source through a 200-micron diameter aperture and a set of metal mesh filters that in combination produce a roughly 20\% wide pass band centered near $\nu=12$~THz ($\lambda = 25$~\micron), as described in Section~\ref{sec:setup}.  

\subsection{Single photon detection}\label{sec:singlephoton}

% \textcolor{red}{Please update this section with new plots and discussion.  This material is just place holder}

\begin{figure}[h]
\centering
\includegraphics[width=.49\textwidth]{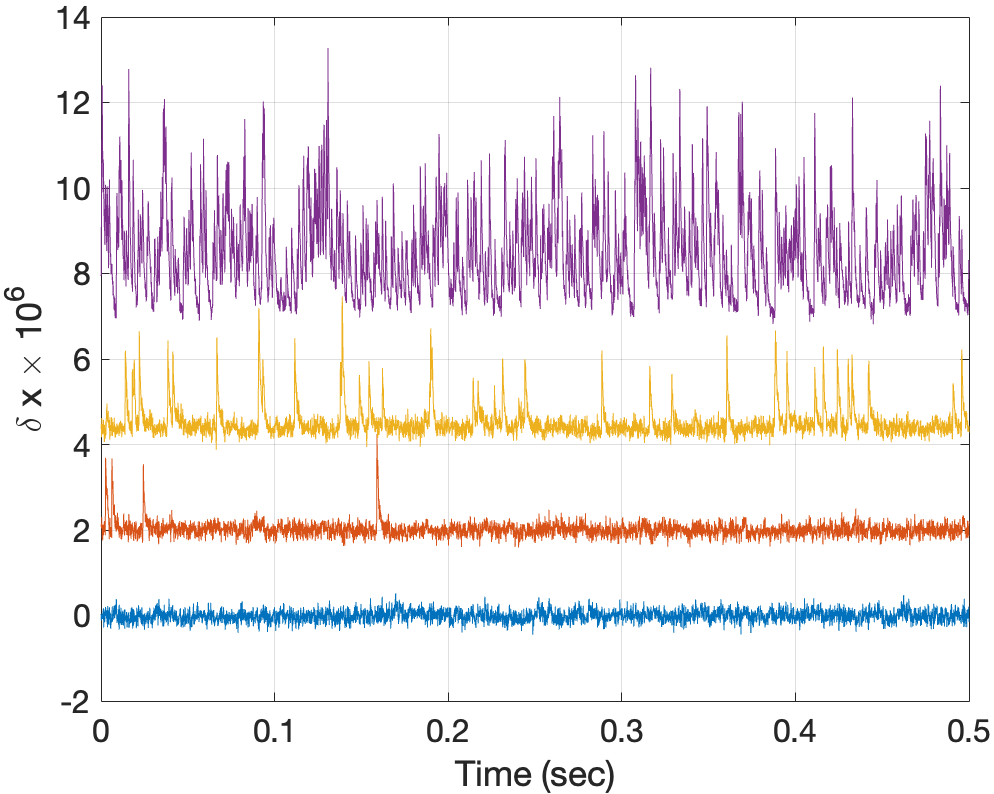}
\includegraphics[width=.47\textwidth]{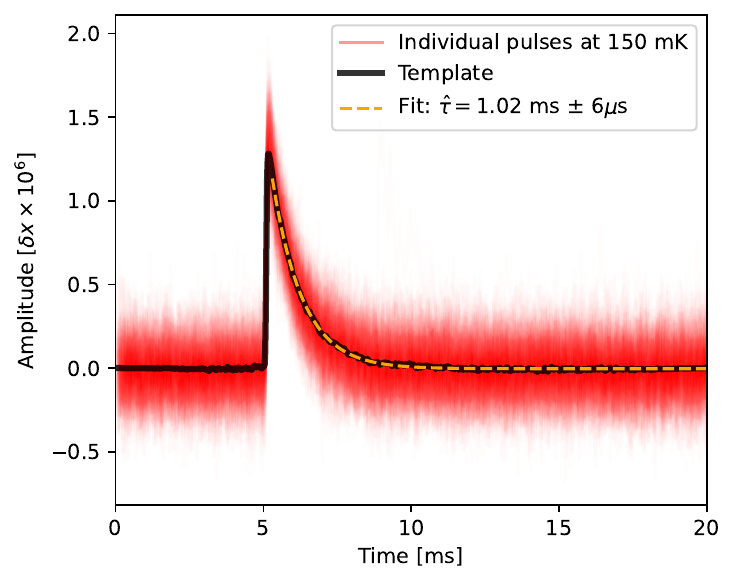} %pulse_stack.pngavepulse.png
\caption{{\it Left:}  Time streams with the black body at (from bottom to top) $T_{BB} = 3$, 30, 35 and 42~K.  The data is offset along the y-axis for clarity.  {\it Right:}  548 stacked single photon pulses (red) at $T_{BB} = 30$K and 50~kHz sample rate.  Alignment is achieved with causal optimal filtering. A single photon template is obtained via averaging (black) and has a time-constant $\approx$ 1~ms, see fit (orange). }
\label{fig:pulse5kHz}
\end{figure}

The filter pass band is in the Wien tail of the black body spectrum for $T_{BB}\sim$30~K and only a few photons per second are expected at the detector. Examples of measured time streams are shown in Fig.~\ref{fig:pulse5kHz} (left) for different black body temperatures.  With the black body at 3K, pulse events are rare.  With increasing temperature high signal to noise pulses are observed with increasing rate and eventually become overlapping. Much larger pulses are also observed with a frequency of about 0.2/s and are attributed to either cosmic ray events or background radioactivity and are excluded from our analyses.  Using a causal optimal filter (OF), aligning and co-adding many pulses of similar amplitude produces a pulse template that is well represented by an exponential decay with a time constant of 1~ms, see Fig.~\ref{fig:pulse5kHz} (right) and Section~\ref{sec:singlephotonmethods}.  The decay time is found to be independent of the pulse height for a particular bath temperature. 

\begin{figure}[h]
\centering
\includegraphics[width=0.9\textwidth]{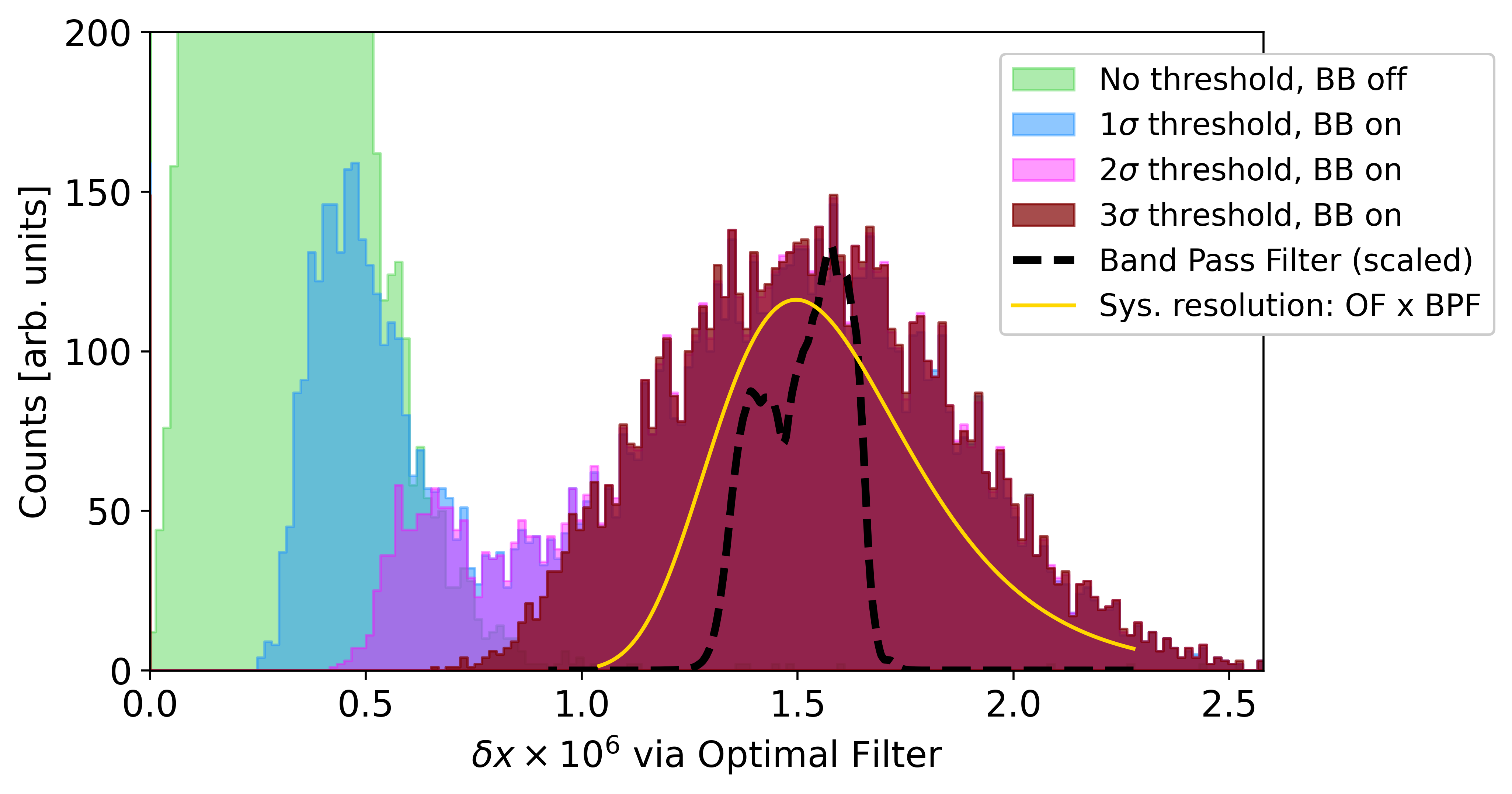}
\caption{Optimal Filter (OF) pulse amplitude histograms. The (black) dashed line indicates the net Band Pass Filter (BPF), i.e., the passband of the metal mesh filters multiplied by the absorber's frequency dependence determined from FTS measurements. The OF energy resolution  convolved with the BPF gives the system resolution (yellow). We run OF on randomly sampled noise data with the black body ``off", without any threshold for pulse selection, and histogram them (green). This is compared with the black body ``on" at 30 K, and varying pulse selection thresholds as effective 1, 2 and 3 $\sigma$ distances from the noise distribution (blue, magenta and maroon, respectively). Thus a clean sample of photon-like pulses are created as the threshold is progressively increased. We particularly note that the system resolution is sufficiently high for Fano-limited single photon measurements.}
\label{fig:pulsehist} %pulsehist32K.png
\end{figure}

A pulse tagging algorithm is used to select photon-like pulses with a threshold criteria as \emph{n} times the standard deviation of the baseline noise. The pulse template then enables extracting the amplitude of each pulse with OF, producing the amplitude histograms shown in Fig.~\ref{fig:pulsehist}, where results are displayed for three threshold values. The histograms show clear Gaussian-shaped peaks with mean $\delta x = 1.54\times 10^{-6}$ which we identify as photons from the black body that pass through the filters.  The excess counts at lower amplitudes \textcolor{black}{include both} %consist of both 
noise-induced false counts and lower-amplitude background photons. The ``3-$\sigma$'' cut effectively rejects all noise spikes and selects the photon pulses, see Section~\ref{sec:singlephotonmethods} for details. 

% Resolving power is quantified as $R = \langle E \rangle/ \Delta E_\mathrm{FWHM}$. An experiment which unambiguously resolves single photons should satisfy the condition that system resolution is higher than the measured resolution, i.e., $R_{sys} > R_{data}$. The system resolution for these measurements is the combined result of the OF energy resolution (limited by readout noise), the product of the measured filter transmission $F(\nu)$ and the frequency dependent absorber efficiency $A(\nu)$. The filter transmission and absorber efficiency were determined from separate FTS measurements, described further in Section~\ref{sec:setup} and Appendix~\ref{sec:res}. For these measurements, $R_{sys} =$ 2.92 and $R_{data}$ is 1.89, where data corresponds to the histogram with the 3$\sigma$ threshold. 

\begin{figure}[h]
\centering
\includegraphics[width=0.7\textwidth]{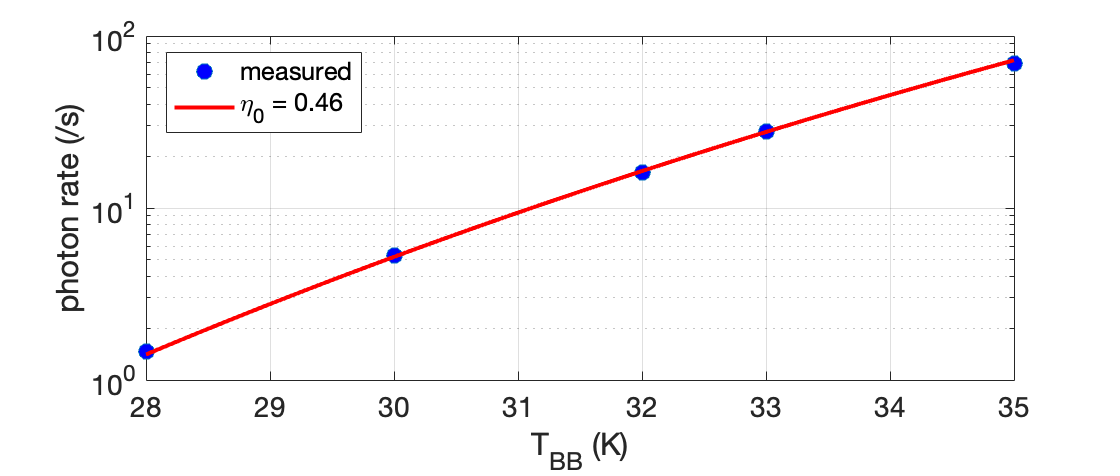}
\caption{Measured and expected photon event rates versus black body temperature. The measured rate (blue dots) was determined by the methods in Section \ref{sec:singlephotonmethods}. The red line represents the expected rate computed via Eq. \ref{eq:photon_rate} where the efficiency factor of $\eta_o=0.46$ what chosen to fit the data, suggesting an unquantified system optical efficiency of this value.}
\label{fig:ratecal}
\end{figure}

The resolving power in energy is defined as $\Rfact = \langle E \rangle/ \Delta E_\mathrm{FWHM}$.   The measured $\Rfact$ necessarily includes all sources of energy broadening.  The two broadening mechanisms in any pulse-detecting instrument are the optical band pass filter (including the wavelength-dependent response of the absorber), and the energy resolution of the detector itself.  The first is determined with FTS measurements of the filter stack transmission $F(\nu)$ and absorber efficiency $\mathcal{A}_{abs}(\nu)$ independently, as described in Section~\ref{sec:setup} and Appendix~\ref{sec:res}.  The detector resolution includes a readout related contribution set by the detector noise floor integrated through the optimal filter and is approximately given by $\mathrm \delta E_{\mathrm rms}\sim NEP\sqrt{\tau}$, where $\tau$ is the pulse decay timescale \cite{Karasik2011ThzPhoton}.  Our estimate of the convolution of these two contributors to $\Rfact$ is shown with the thin yellow curve in Fig.~\ref{fig:ratecal}.  We compute an effective resolving power is 2.92, compared to this, the data has an $\mathcal{R} =$ 1.89. 

Our pair-breaking detector is also subject to statistical variation in the finite number of quasiparticles generated and their timescales for recombination. This Fano limit is given by:
\begin{equation}
    \Rfact_\mathrm{FANO} = \frac{1}{2\sqrt{2\log{2}}} \sqrt{ \frac{\eta_{pb} E_\gamma}{\Delta F}} \approx 8,
\end{equation}
where the value of 8 is what would be expected with a Fano factor F~$\simeq$~0.2 \cite{kurakado1982possibility}, and the pair breaking efficiency of 0.3 estimated below.  This perfect Fano limit should thus not limit the $\Rfact$ in our system.
The fact that the measured resolving power ($\Rfact_{meas}\sim 1.9$) is lower than the expected value, combined with the relatively low pair breaking efficiency suggests  that energetic phonon loss \cite{kozorezov2007electron} is a contributing element to the response of the detector. This is not surprising as the photon energy is similar to the Debye energy, so the phonon-controlled phase of the energy cascade process is similar to that which occurs in detectors for visible and higher energy photons where energetic phonon loss clearly occurs \cite{de2021phonon, zobrist2022membraneless, kozorezov2008phonon}. 

The measured photon arrival rate $\Gamma_\gamma$ for black body with temperature $T$ is determined from the area under the Gaussian fit to the pulse amplitude histogram (Fig.~\ref{fig:pulsehist}) in order to exclude noise-induced false counts and low-energy background events.  $\Gamma_\gamma(T)$ is shown in Fig.~\ref{fig:ratecal} for black body temperatures low enough that single photons are observed without significant pile up.  The measured rates are compared to the integrated black body spectral photon radiance, $B(\nu, T)/h\nu$, multiplied by  $F(\nu)\mathcal{A}_{abs}(\nu)$, and constant efficiencies related to the optical system:
\begin{equation}
\Gamma_\gamma(T) = \eta_o \epsilon A \Omega \int \frac{B(\nu, T)}{h\nu} F(\nu) \mathcal{A}_{abs}(\nu) d\nu,
\label{eq:photon_rate}
\end{equation}
where $A$ is the microlens area, based on the measured 870~\micron\, diameter, $\Omega$ is the solid angle subtended by the 200~\micron\, diameter aperture, and $\epsilon$ is the effective emissivity of the black body including a small reduction in coupling to the aperture due to diffraction \cite{blevin1970diffraction}.  The parameter $\eta_o$ accumulates the efficiency with which the microlens couples incident power onto the absorber along with unquantified efficiencies in the system such as an aperture thickness effect.  We find that $\eta_o = 0.46$ matches the data.  Much of this loss is likely to be due to a deviation in the fabricated lens profile from the design around the edges of the lens.  A diffraction calculation using the measured microlens profile suggests a spill-over efficiency of 0.70, compared to 0.94 for the designed profile.  The fabricated profile also results in a more tightly focused optical spot on the absorber than intended, which reduces the coupling relative to the infinite array behavior, as the spot size becomes comparable to the size of the unit cell of the absorber pattern.  Importing the field distribution from the diffraction calculation into a full-wave EM simulator, we estimate a coupling efficiency of 0.87 compared to optimal defocusing.  Further losses of a few percent are expected from the antireflection layer on the lenses and from absorption in the epoxy bond between the detector and microlens array wafers.

\subsection{Quasiparticle conversion efficiency from pulse height}

Photon absorption in a superconducting film initiates a phonon mediated cascade process whereby the energy of the initial photoelectron is transferred to a population of quasiparticles near the gap energy with  efficiency $\eta_{pb} = \Delta_0 N_{qp} / h\nu$, where $h\nu$ is the absorbed photon energy and $N_{qp}$ is the number of gap edge quasiparticles produced, and $\Delta_0$ is the superconducting gap energy.  The remaining energy is lost to phonons.  In a thick superconducting film, these phonons have energy $\Omega < 2\Delta_0$ so that they are unable to break Cooper pairs.  In this case a calculation has found $\eta_{pb} \approx 0.6$ for aluminum \cite{kozorezov2000quasiparticle}.  A similar efficiency has been found for thin films at low photon energies (e.g. 0.85~THz) where only a few quasiparticles are created per photon, based on comparing photon noise with thermal generation-recombination noise \cite{Janssen_22}.  On the other hand at optical and higher photon energies, energy loss due to escape of $\Omega > 2\Delta_0$ phonons into a solid substrate has been found to be a major limitation on the energy resolution of thin film MKID and STJ detectors, as the loss of these energetic phonons increases the statistical fluctuation in the detected energy.   \textcolor{black}{For example, our measurement of 1.55~$\mu m$ (193~THz) photon pulses in our aluminum KIDs suggests a pair breaking efficiency of 0.13 at these near-IR energies \cite{Kane_24}. }

The efficiency for quasiparticle production by photons with average frequency $\langle \nu \rangle = 12.1$~THz in the 40~nm aluminum film of our detector can be estimated from the average measured pulse height.  Using the value of $\delta x / \delta N_{qp}$ determined in appendix \ref{sec:dxdNqp},
\begin{equation}
\eta_{pb} = \frac{\Delta_0}{h \langle\nu\rangle} \frac{\langle\delta x\rangle}{(\delta x / \delta N_{qp})} = 0.32.
\end{equation}
This value agrees well with the value found from the measured noise spectra described below.  \textcolor{black}{It is unsurprisingly intermediate between the low-energy limit of $\sim$ 0.6 and the much lower values at optical / near-IR energies.  It also agrees with a theoretical treatments of the energy dependence of the pair-breaking efficiency \cite{kozorezov2008phonon,guruswamy2014quasiparticle}.}

\subsection{Photon-shot-noise-limited sensitivity}

\begin{figure}[h]
\centering
\includegraphics[width=1.00\textwidth]{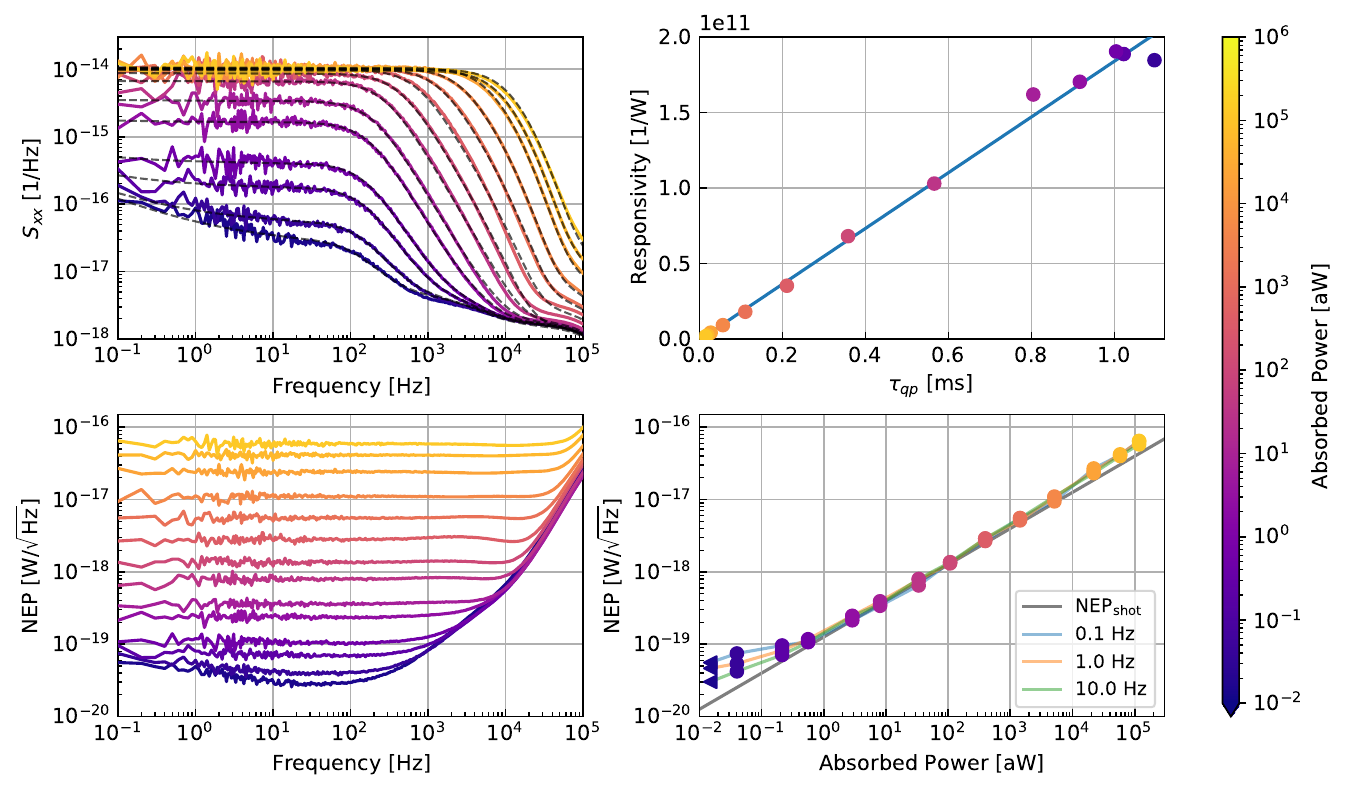}
\caption{Detector noise, time constants, and sensitivity as a function of absorbed optical power, which includes the optical efficiency factor of 0.46 determined from the photon count rate in Section~\ref{sec:singlephoton}. Upper left: fractional frequency noise spectra show for multiple absorbed powers. Black dashed lines show noise spectra fits incorporating a white noise component, the quasiparticle lifetime, the resonator ringdown time, and a TLS component (only significant for the lowest absorbed powers). Upper right: The measured responsivity as a function of the quasiparticle lifetime extracted from the fractional frequency noise spectra fits, including a linear fit overplotted. Lower panels: NEP as a function of audio frequency (lower left) and absorbed power (right). Data points with the $\blacktriangleleft$ symbol illustrate measurements with the black body infrared source turned off where the true absorbed power of these data points is many orders of magnitude lower. The NEP is shot noise limited over nearly six orders of magnitude of absorbed power.}
\label{fig:NEPcombo}
\end{figure}

The upper left panel of Fig.~\ref{fig:NEPcombo} shows the fractional frequency noise power spectral densities (PSDs) for a range of optical powers. The noise spectra are well characterized by a low frequency white component and a roll off, which contains information on the decay time of the quasiparticles, $\tau_{qp}$. At lower optical powers, a low frequency component is also present that scales approximately as $f^{-1/4}$ in the range shown, indicative of TLS noise. Fits to the noise spectra are overlaid with black dashed lines. The best fit $\tau_{qp}$ values are plotted in the upper right panel of Fig.~\ref{fig:NEPcombo} against the measured fractional frequency optical responsivities for each optical power. The responsivity measurement method is described in Section~\ref{sec:measurement_responsivity}. $\tau_{qp}$ and the responsivity scale linearly with each other, as illustrated with the best fit line.

The lower left panel of Fig.~\ref{fig:NEPcombo} shows the noise equivalent power (NEP) as a function of audio frequency for a range of optical powers. The low-frequency stability of the detector results in stable NEP performance over a wide range of audio frequencies for all optical powers. The lower right panel of Fig.~\ref{fig:NEPcombo} shows the NEP at 0.1, 1, and 10 Hz as a function of absorbed power. The absorbed power calculation assumes includes the optical efficiency determined from the single photon arrival rates described in Section~\ref{sec:singlephoton}, $\eta_o = 0.46$. When this efficiency is included, the NEP data fall on the theoretical photon shot-noise curve given by $NEP_{shot}=\sqrt{2h\nu P_{abs}}$ over six orders of magnitude of absorbed power.
At the lowest absorbed power, where the detector can sense single photon events, we find a limiting NEP of 
$4.6\times10^{-20}~{\rm W/\sqrt{Hz}}$ at 1~Hz for an integrating readout.  The NEP in the photon counting regime is ultimately limited by the dark count rate.

% Measured fractional frequency noise power spectral densities (PSDs) are shown for a range of optical powers in the upper left panel of Fig.~\ref{fig:NEPcombo}.  The noise spectra are well characterized by a low frequency white component and a roll off, which contains information on the decay time of the quasiparticles, $\tau_{qp}$.  

The increase in the white component of the fractional frequency PSDs with power at low optical power and the saturation at high power has been observed in previous studies of MKIDs and has been associated with the saturation of $\tau_{qp}$ \cite{de2014fluctuations}.  As quasiparticle recombination is a pairwise process, the lifetime decreases with the quasiparticle density $n_{qp} = N_{qp} / V$ but reaches a maximum value, so that
\begin{equation}
\tau_{qp}^{-1} = 2 R n_{qp} + \tau_{max}^{-1}.
\end{equation}
where $R$ is the recombination constant. The mechanism by which $\tau_{qp}$ saturates is not understood, but it has been suggested it results from quasiparticle trapping.  

The reduction in $\tau_{qp}$ with $n_{qp}$ results in an optical response that decreases with increasing power absorbed by the detector, $P_{abs}$, as
\begin{equation}
\frac{\delta n_{qp}}{\delta P_{abs}} = \frac{\eta_{pb}\tau_{qp}}{V \Delta_0}  = \frac{\eta_{pb} \tau_{max}}{\Delta_0 V}\frac{1}{\sqrt{1 + \eta_{pb} P_{abs} / P_\ast}},
\end{equation}
with $P_\ast = \Delta_0 V / (4 \tau^2_{max} R)$.  Under optical illumination, the photon shot noise contribution to the PSD of fractional frequency fluctuations is then $S_{xx} = S_N (\delta x / V\delta n_{qp})^2$, where the quasiparticle number fluctuation is 
\begin{equation}
S_N^\gamma(f) = 2h\nu P_{abs} (\frac{V\delta n_{qp}}{\delta P_{abs}})^2  = \frac{\eta_{pb} h\nu V}{2 \Delta_0 R}\frac{\eta_{pb} P_{abs} / P_\ast}{1 + \eta_{pb} P_{abs} / P_\ast}  \frac{1}{1 + (2\pi\tau_{qp}f)^2}, 
\label{eqn:Sxphoton}
\end{equation}
which has a roll-off determined by $\tau_{qp}$.  The wave noise contribution is neglected as the photon occupation number $n_0 \ll 1$.  The white noise region of the spectrum approaches a constant value 
%$S^\gamma_N \longrightarrow {\eta_{pb} h\nu V} / {\Delta R}$ 
in the limit of high optical power $P \gg P_\ast$.  Pair breaking detectors incur an additional noise term due to the recombination of quasiparticles.  Expressed as a fractional frequency noise  
$S^{
R}_{N} = 2 N_{qp} \tau_{qp} / (1 + (2\pi\tau_{qp}f)^2)$ \cite{wilson2004quasiparticle}, 
which approaches $V/R$ in the high optical power limit.  The recombination noise is much smaller than the photon noise as $\eta_{pb} h\nu \gg 2\Delta_0$.

% \begin{sidewaysfigure}[h!]
% \includegraphics{figures/NEPcombo.png}
% \caption{Fractional frequency noise $S_{xx}$ for estimated incident optical powers, from blue to red, 0~zW, .... }
% \label{fig:NEPcombo}
% \end{sidewaysfigure}

% \begin{figure}[h]
% \centering
% \includegraphics[width=1.00\textwidth]{figures/NEP.eps}
% \caption{NEP referred to power incident on the lenslet (colored lines) for estimated optical powers, from blue to red, 0~zW, ....  The dashed line represents the NEP with respect to absorbed power.}
% \label{fig:NEP}
% \end{figure}

% \begin{figure}[h]
% \centering
% \includegraphics[width=1.00\textwidth]{figures/Sdff_0119-0304_602MHz_paper.pdf}
% \caption{Caption text goes here.}
% \label{fig:Spar_602MHz}
% \end{figure}

\subsection{Quasiparticle conversion efficiency from noise}
\label{sec:etaqpnoise}

A second method of deriving $\eta_{pb}$ compares the measured noise under photon illumination to thermal generation-recombination (GR) noise and provides $\eta_{pb}$ independently of knowledge of both the precise inductor volume and the quasiparticle response $dx / dN_{qp}$.   

The GR noise was measured by raising the base temperature of the detector.  The noise spectra are shown in Section~\ref{sec:basetresp}, and show a similar behavior to the noise spectra under optical illumination.  The low frequency white noise component of the Lorentzian spectra increases initially with increasing temperature and then saturates when the quasiparticle density becomes high enough that $R n_{qp} \gg 1/\tau_{max}$.  At that point, $S^{GR}_N(0) = 4 N_{qp}\tau_{qp} = 2V/R $ \cite{wilson2004quasiparticle}.  Comparing the measured photon and thermal GR noise saturation values, we find that
\begin{equation}
\frac{S_x^\gamma(0) + S_x^{R}(0)}{S_x^{GR}(0)} = \frac{1}{2}( \frac{\eta_{pb} h \nu}{2\Delta_0} + 1) = 17.6,     
\end{equation}
which suggests $\eta_{pb} = 0.28$, in rough agreement with the single photon pulse height analysis presented above. Using this value in Equation~\ref{eqn:Sxphoton} and comparing to the measured photon noise saturation with $\delta x / \delta N_{qp} = 2\times 10^{-8}$, we find the recombination constant $R = 14.8\,\mu\mathrm{m}^3/\mathrm{s}$.

\subsection{Integrating mode dynamic range}

From Fig.~\ref{fig:NEPcombo} it can be seen that the detector is near shot noise limited over about six orders of magnitude in absorbed power.  Because the response $\delta x / \delta P$ of a KID falls off with increasing optical power as $P^{-1/2}$ for $P > P_\ast$ while the sensitivity required to be photon noise limited is reduced by the same factor, the dynamic range of the detector is potentially quite large.  A practical limit to the power that can be absorbed while maintaining shot noise limited sensitivity is imposed by the degradation of the quality factor due to quasiparticle loss.  The loss per quasiparticle is roughly constant and is related to the frequency response as
\begin{equation}
\frac{Q_i^{-1}}{N_{qp}} = \frac{2 \sigma_1(f_r, T)}{\sigma_2(f_r, T)} \frac{\delta x}{\delta N_{qp}} ,    
\end{equation}
where $\sigma(f_r, T) = \sigma_1 - i\sigma_2$ is the complex conductivity. At 600~MHz, $\sigma_1 / \sigma_2 \approx 0.15$, several times less than at GHz frequencies typically used for KID readout.  Hence the response of the resonator loss is reduced relative to the frequency response for low readout frequencies, enhancing the dynamic range.  The lowest $Q_i$ that can be tolerated is primarily determined by the multiplexing factor.  For 1000 detectors per readout line, it is reasonable to require $Q_i \gtrsim 10^4$.  Using $N_{qp} \approx (V\eta_{pb}P_{abs}/R\Delta)^{1/2}$, we estimate $Q_i \approx 10^4$ with $P_{abs} = 100$~fW, in good agreement with the measurements.

\subsection{Low frequency stability}
% \begin{figure}[h!!]
% \centering
% \includegraphics[width=0.9\textwidth]{figures/Allan_watts.png}
% \caption{Allan--deviation based minimum detectable power ($P_\sigma$) w.r.t varying integration times ($\tau$) and optical loading (colored lines/ patches encompassing statistical errors). White-noise scaling  (dashed black line) and typical frequency modulation range for FIR / MIR telescopes using MKIDs (gray block) are shown for guidance. As seen here, at powers $\gtrsim 1$ aW, the detector is stable for $\gtrsim 500$ s. }
% \label{fig:allan}
% \end{figure}
% \begin{figure}[h!]
% \centering
% \includegraphics[width=.9\textwidth]{figures/lowfPSD.png}
% \includegraphics[width=.9\textwidth]{figures/Allan_watts.png} %avepulse.png
% \caption{{\it Top:}  Amplitude Spectral Density of noise at varying black body optical loads, note the flat ASD down to mHz for loading above a few aW.  {\it Bottom:}  Allan--deviation based minimum detectable power ($P_\sigma$) w.r.t varying integration times ($\tau$) and optical loading (colored lines/ patches encompassing statistical errors). White-noise scaling  (dashed black line) and typical frequency modulation range for FIR / MIR telescopes using MKIDs (gray block) are shown for guidance. As seen here, at powers $\gtrsim 1$ aW, the detector is stable for $\gtrsim 500$ s. \ifcomment\textcolor{red}{This plot is not scaled correctly - should correspond to the 1Hz NEP at 1s integration time}\fi }
% \label{fig:allan}
% \end{figure}
\begin{figure}[h!]
\centering
\includegraphics[width=0.49\textwidth]{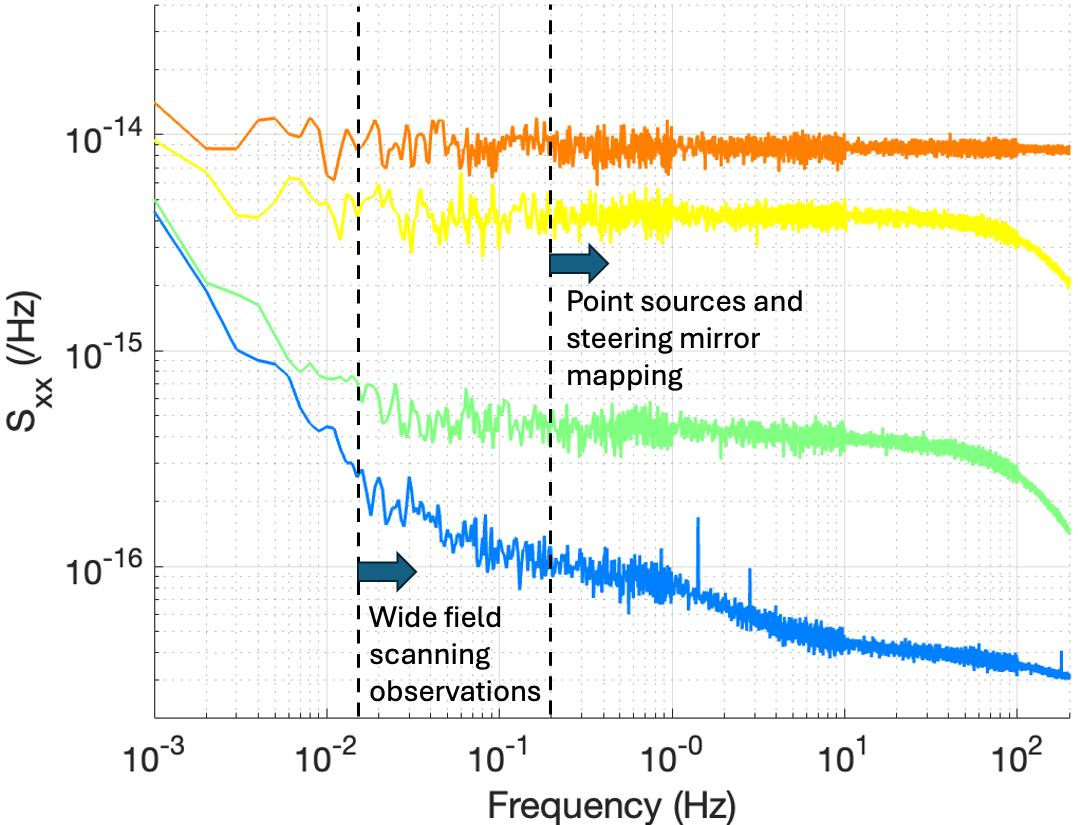}
\includegraphics[width=0.49\textwidth]{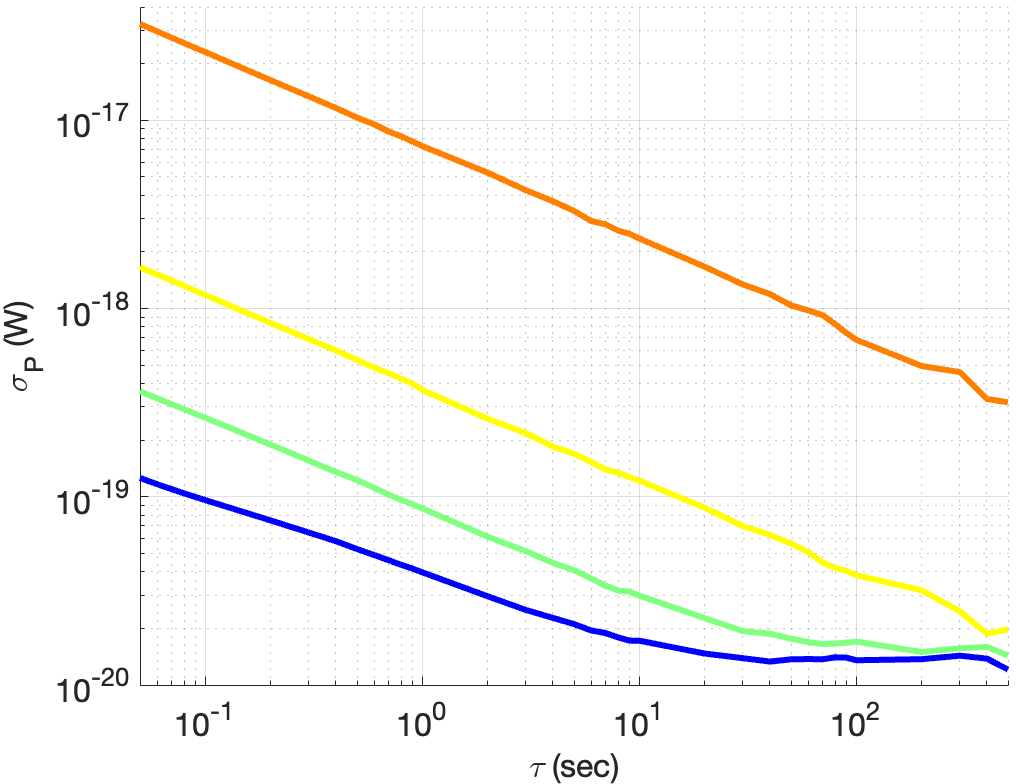}
\caption{{\it Left:} Fractional frequency noise power spectra for (from bottom to top) $P_{abs} =$ 0, 0.6~aW, 11~aW and 3.5~fW.  {\it Right:}  Corresponding Allan deviation plot showing stability out to tens to hundreds of seconds.}
\label{fig:allan}
\end{figure}

We have studied the stability of our devices by obtaining 6 hour datasets with the detector held at 150 mK and the blackbody controlled at fixed temperature.  The data are presented both as power spectra and as the Allan deviation in Figure~\ref{fig:allan}.  We find these detectors to be stable to $\sim$100 seconds at $<$ 0.1 aW loading and to $>$ 300s for loading $>$ 7 aW, corresponding to frequencies of 10 mHz and 3 mHz, respectively.  The dependence of the noise knee frequency, or corresponding Allan time inflection point on loading is expected, since the white photon noise increasingly overtakes the low-frequency device noise as loading is raised.  The results are consistent with an additive frequency noise term that may be from two level systems.

%Allan--deviation equivalent measurements were made to determine %the minimum detectable power $P_\sigma$ versus integration time %$\tau$ for varying optical power levels by collecting long time %streams with the black body controlled at fixed temperature. %The detector held at 150~mK.  We find these detectors to be %stable to $\sim$100 seconds at $<$ 0.1 aW loading and to $>$ %300s for loading $>$ 7 aW, Fig.~\ref{fig:allan}. Thus the %minimum detectable power can be lowered by integrating longer, %to hundreds of seconds or longer, before inflections due to low %frequency noise. The inflection point has an expected %dependence on the optical loading, i.e., for higher loading the %inflection is at longer time scales. 

Modulation frequencies of $\sim$0.5 to $\sim$350 Hz have been envisioned for most science experiments with mid- and far-IR space missions thus far, in part due to limitations of previous detector systems.  Our demonstrated low-frequency performance exceeds these requirements, and offers additional opportunities.
%The modulation frequency range for mid / far infrared telescopes using MKIDs is usually below 0.2 seconds for comparison. 
%This demonstrated low frequency stability opens up new 
%observational scopes beyond what was considered for the proposed FIR studies.  
It both relaxes requirements on the modulation mechanisms (which are necessarily cryogenic), and enables unique science cases which require long temporal stability. 
One example is wide-field spectral mapping, also known as line intensity mapping, which requires recovering large-scale (many-degree) modes of the Universe \cite{Karkare_22}; this necessitates spacecraft scanning which generates modulation frequencies as low as 15 mHz.
Another is studying planetary atmospheres as they transit in front of and behind their hosts stars; these timescales are on order hours \cite{Bouwman_2023}, corresponding to sub-mHz frequencies, but typically are accompanied by a large photon noise. %... \textcolor{red}{Can we be specific here?}\fi

\begin{figure}[h!]
\centering
\includegraphics[width=0.8\textwidth]{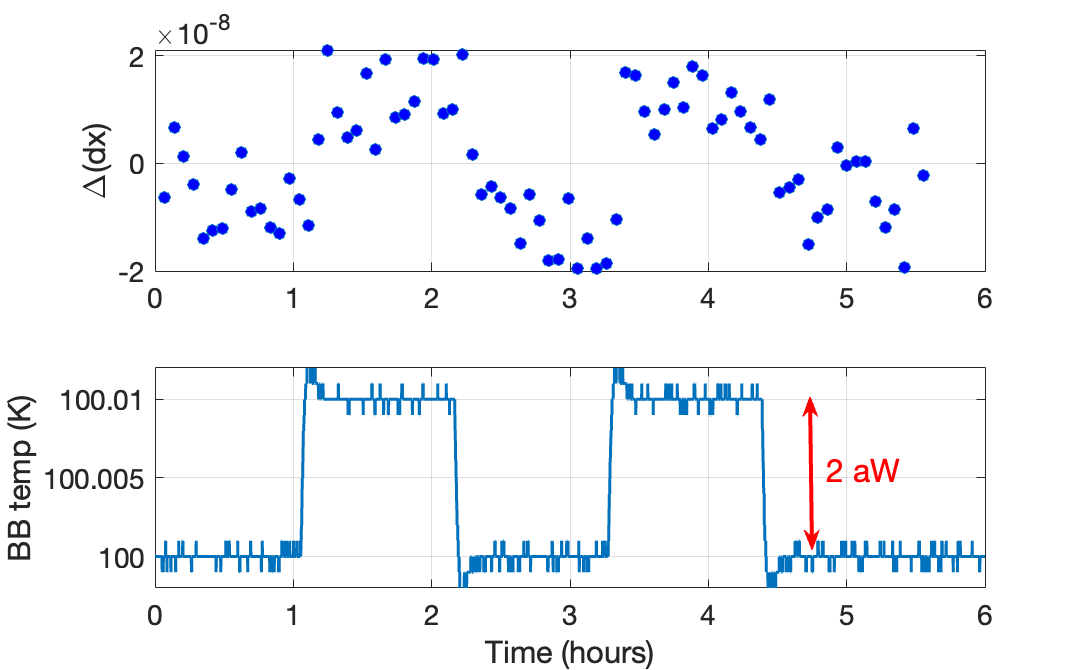}
\caption{Time stream data (top panel) taken with the black body modulated between 100 and 100.01~K with a period of 8000~seconds.  A linear drift term was removed.  A 16\% transmission neutral density filter was inserted between black body and detector for this measurement.  The average absorbed power was 3.5~fW and the modulation was 2~aW.  The black body thermometer record is shown in the bottom plot.}
\label{fig:mod100K}
\end{figure}

A demonstration of the excellent stability under load is shown in Figure~\ref{fig:mod100K}.  Here we show a portion of a time stream taken under a relatively high optical loading of 3.5~fW.  The black body was modulated by changing its temperature by 10~mK with a period of 8000~seconds.  The points shown are averages over 250 seconds and a linear drift of $\delta x / \delta t = 8\times 10^{-9}$ per hour was removed.  The shot noise limit for a 250 second integration time corresponds to $\delta P$(rms) $= 4.9\times 10^{-19}$~W, which is close to the measured noise of $6.6\times 10^{-19}$~W, verifying that the noise integrates as expected under optical loading.

\section{Methods}\label{sec11}

\subsection{Optical Setup} \label{sec:setup}

\begin{figure}[h]
\centering
\includegraphics[width=0.5\textwidth]{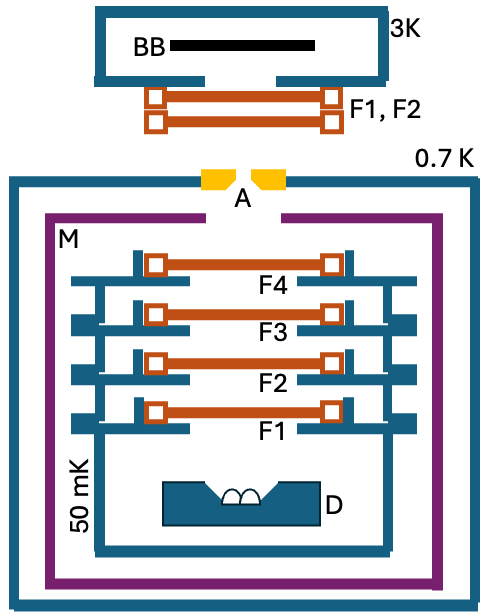}
\caption{Diagram of the optical setup. A temperature variable black body (BB) is anchored to the 3K stage of the dilution refrigerator system and can be heated to $\approx$ 140~K. The microlens-hybridized detector chip (D) is mounted on the 50~mK stage of the refrigerator and is surrounded by a cryoperm magnetic shield (M).  It is exposed to the black body through a 200~\micron\, diameter aperture and a set of metal mesh filters:  F1 ($400~\rm{cm}^{-1}$ bandpass) and F2 ($300~\rm{cm}^{-1}$ highpass) together form the pass band for the black body and F3 ($600~\rm{cm}^{-1}$ lowpass) and F4 ($900~\rm{cm}^{-1}$ lowpass) provide protection from short wavelength optical leakage.}
\label{fig:opticalsetup}
\end{figure}

The detector array is mounted to the 50~mK stage of a dilution refrigerator inside a housing designed with the goal of preventing leakage of background thermal photons from higher temperature stages of the cryostat. The cryogenic and optical setup is illustrated in Fig.~\ref{fig:opticalsetup}. A set of 4 metal mesh filters, manufactured at Cardiff University, limit radiation into the detector housing.  Two of these filters (F1 and F2) are bandpass and highpass filters, which together form a band pass around $\nu = 12$~THz ($\lambda=25$~\micron).  Filters F3 and F4 are low pass filters used to block thermal photons that might leak into the system through the pump line. The transmission of all four filters was measured cryogenically in an FTS by the Optics Branch at Goddard Space Flight Center. The detector housing is surrounded by a cryoperm magnetic shield.  These structures are located within the 0.7~K thermal shield of the refrigerator, which has been carefully sealed to prevent optical leaks except for a 200~\micron\ diameter aperture in a 50~\micron\ thick copper foil through which the detectors view the black body.

The black body source is a 25~mm diameter aluminum disk with a surface coated with Aktar Metal Velvet and is located in a separate housing mounted to the 3~K stage of the refrigerator.  The emissivity of the Metal Velvet surface was measured at room temperature at JPL and found to be 0.92 within the bandpass of the Cardiff filters.  The source is thermally isolated with a Kevlar thread suspension system and can be warmed with a resistive heater to at least 140~K without significantly perturbing the cooling system.  An additional set of the F1 and F2 filters limit the optical band emitted from the black body housing.  Due to the small size of the aperture compared to a wavelength, it is necessary to include a diffraction correction \cite{blevin1970diffraction} that characterizes the coupling of the aperture to the black body, which we estimate to be 0.9.

\subsection{Measurement of $\delta x$} \label{sec:deltax}

\begin{figure}[h!]
\centering
\includegraphics[width=1\textwidth]{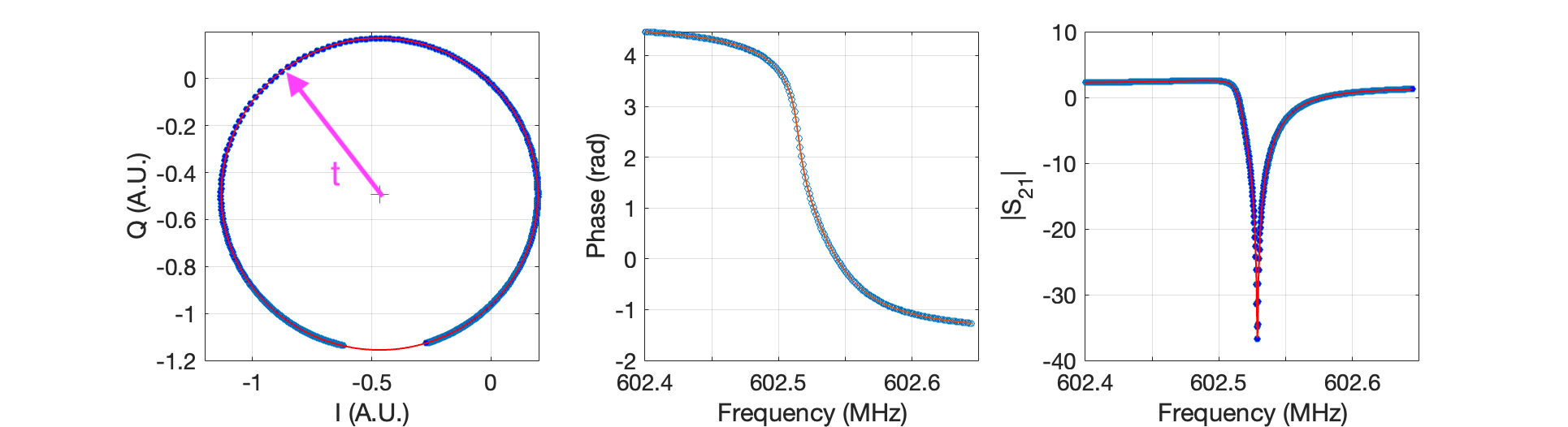}
\caption{{\it Left:} Resonance loop corrected for cable delay.  }
\label{fig:IQ}
\end{figure}

Prior to measuring a particular detector, its resonance curve is determined by sweeping the synthesizer frequency and recording the $I$ and $Q$ outputs of the mixer (Fig.~\ref{fig:IQ} left).  The complex data is parameterized in terms of a single phase value, $\theta$, representing the angle of the $(I,Q)$ point with respect to the center of the resonance loop.  The $\theta$ versus frequency curve is fit to a nonlinear resonator model \cite{swenson2013operation, dai2022new} (Fig.~\ref{fig:IQ} center and right).  The $\theta(f)$ result from the fit is then inverted and used to calibrate measured I and Q time streams into units of fractional frequency shift, $\delta x = \frac{f-f_0}{f_0}$.

\subsection{Measurement of $\delta x / \delta P_{abs}$}
\label{sec:measurement_responsivity}

\begin{figure}[h!]
\centering
\includegraphics[width=.8\textwidth]{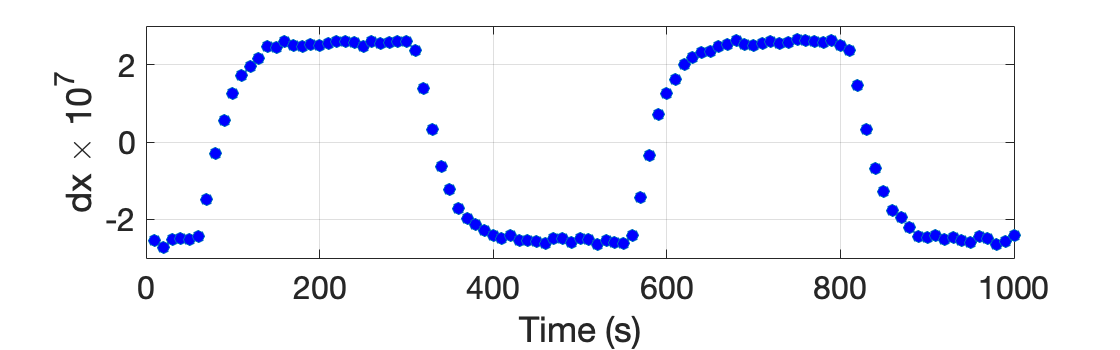}
\caption{Average of two periods of the black body temperature modulation between 80~K and 80.12~K. }
\label{fig:dxdp}
\end{figure}

% The response of $\delta x$ to optical power at different optical power levels 
The fractional frequency responsivity as a function of optical power, $R_{\delta x}(P_{abs} = \frac{{\rm d}\delta x}{{\rm d}P_{abs}}(P_{abs})$,  was determined by imposing a small modulation on the black body temperature.  The modulation amplitude was chosen to produce an optical power modulation of the greater of either 1~aW or one percent of the average power.  Because of the long thermal time constant of the black body, the dwell time at each temperature was set at 250~s and a time stream of several hours was collected and averaged over the modulation period, as shown in Fig.~\ref{fig:dxdp}. The frequency shift for each period is determined by differencing the average high- and low-amplitude sections of the modulation. Data during the modulation transition time are ignored. The responsivity is taken as the mean frequency shift over the course of the several hour modulated time stream. 

\subsection{Single-photon data analysis method} 
\label{sec:singlephotonmethods}

A pulse selection algorithm is used to generate clean samples of single photon pulses from 50 kHz 
\textcolor{black}{fractional frequency shift}
time streams. This is done by (i) applying an order 3 median filter,  (ii) convolving with a simple pulse model (falling exponential) to identify temporal locations of potential pulses, (iii) tagging events that stay above an amplitude threshold for a certain number of samples, (iv) removing tagged events above a second, much-higher amplitude threshold (to eliminate cosmic ray events), leaving the remainder as single photon candidates, and (v) selecting on a rise time criterion characteristic of energy deposition processes. The threshold value is given in multiples of the standard deviation ($\sigma$) of the total signal with the black body at 30~K. We find 3-$\sigma$ as the optimal threshold for detecting 25~\micron\ photons and rejecting noise spikes. An amplitude cutoff value can be set as an additional selection measure to isolate the noise peak in Fig.~\ref{fig:pulsehist}, where a $\delta x > 0.9\times10^{-6}$ cutoff retains 97\% of the selected pulses. These successive steps provide time-stamps used to pick out photon-like pulses from the raw time streams. The photon-like pulses are then processed with an Optimal Filter routine (OF) \cite{basu_thakur_cryogenic_2015}. Since the OF only evaluates individual pulses, pulses separated by less than 1.6 ms from a temporal neighbor are discarded. We perform a first-pass causal OF with a simple template to extract pulse onset times % used to determined the
\textcolor{black}{and then stack and average pulse events to construct a}
single photon template, Fig.~\ref{fig:pulse5kHz}. This template is then used for a second-pass causal OF. We perform basic data-quality $\chi^2$ and timing cuts to select a batch of clean single-photon events which are studied in detail, Fig.~\ref{fig:pulsehist}.

The dark count rate (DCR) is computed by applying the pulse selection algorithm to time streams collected with the black body at 3~K. The algorithm is set to the 30~K 3-$\sigma$ threshold (clean selection of 12 THz photons in the optical measurements, maroon histogram in Fig.~\ref{fig:pulsehist}). Cosmic ray events are excluded from the DCR in two steps. The pulse selection algorithm automatically separates pulses above a maximum threshold amplitude. To account for cosmic rays that impact the chip far from the studied resonator and thus are seen as low-amplitude pulses similar to 25~\micron\ photons, we use a coincidence trigger between the 602 MHz resonator and a neighboring pixel to flag simultaneous events within a 40~$\mu$s window.

The DCR for pulse-like events with amplitudes
$\delta x > 1.2 \times 10^{-6}$ is 4.4 $\text{mHz}$. As this amplitude cut-off is lowered, more pulse-like events are captured leading to an increased DCR. Tab.~\ref{tab:dcrtable} shows some representative values and we note lower detection efficiencies for lower pulse amplitudes, i.e., when SNR is lower.
These efficiencies are computed through data-driven simulations. Noise is simulated from 3 K black body time streams with random phase rotations which maintain the 
measured fractional frequency
PSD. A comb of pulses, derived from the OF template with varying amplitudes, are added to the synthetic noise. The amplitude dependant efficiency is computed as the passage fraction of the comb detected by the pulse-selection algorithm, weighted by the occurrence of the comb amplitudes in the measured distribution. %In general, such efficiency profiles acting on the noise distribution define the lower amplitude peaks in Fig.~\ref{fig:pulsehist}.  

\begin{table}[h]
\centering
\begin{tabular}{|l|l|l|l|}
\hline
Cutoff Amplitude ($\delta x$)    & 0.9 $\times 10^{-6}$  & 1.0$\times 10^{-6}$   & 1.2$\times 10^{-6}$    \\
\hline
Dark Count Rate      & 6.6$\text{ mHz}$ & 5.3$\text{ mHz}$ & 4.4$\text{ mHz}$  \\
\hline
30~K Population Above Cutoff & 97\% &  94\% & 80\% \\
\hline
30~K Detection Efficiency & 93\%   & 96\%   & 99\%   \\
\hline
\end{tabular}
\caption{The effects of minimal cutoff amplitudes on
the DCR and the 30~K photon distribution. The DCR was computed with 8000~seconds of data, with the black body at 3~K. Both the 3~K DCR and 30~K distribution were processed through our 3-$\sigma$ pulse selection algorithm and corrected for the dead time of the algorithm. }
\label{tab:dcrtable}
\end{table}

The dark count rates in table~\ref{tab:dcrtable} reflect monitoring of two detectors simultaneously to allow some removal of coincident events, described in Appendix \ref{sec:ctrigger}.  The remaining pulses largely have photon-like rise and decay times and may be from thermal photons leaking into the cryostat through the pump line.  The ultimate limits to the DCR will be evaluated in future measurements using a fully dark device housing and a full-array multichannel readout system to allow more reliable simultaneous event detection.

% simpler version above in red, more metrics here if needed.
% Dark count rate (DCR) is computed by applying the pulse selection algorithm to time streams collected with the black body at 3~K. Applying the 3-$\sigma$ threshold, DCR for pulse--amplitudes $\delta x /10^{-6} >$ 0.98, 1.2, and 1.4, are 0.025, 0.018 and 0.012 $\text{s}^{-1}$ respectively (sampling 2000s of data). These amplitudes correspond to pulse selection algorithm's efficiencies (for the 3-$\sigma$ criteria) of $>$50\%, 90\% and 99\%, respectively. These efficiencies are computed through data-driven simulations. Noise is simulated from 3 K black body time streams with random phase rotations which maintain the power spectral density. A comb of pulses, derived from the OF template with varying amplitudes, are added to the synthetic noise. Amplitude dependant efficiency is computed as the passage fraction of the comb detected by the pulse-selection algorithm. In general, such efficiency profiles along with the noise distribution define the lower amplitude peaks in Fig.~\ref{fig:pulsehist}. 

\section{Discussion}\label{sec:discussion}

The measurements that we have presented describe a very versatile detector that is enabling for applications over a wide range of input optical powers.  Very low power applications include direct dark matter detection experiments or space-borne far- to mid-IR spectrometer or interferometer instruments operating near the background limit set by zodiacal dust emission.  \textcolor{black}{One leading example is the FIRESS spectrometer \cite{Bradford2024FIRESS} on the PRIMA far-IR PRobe \cite{Glenn_2024}, for which these detectors are prototypes. In these low-power applications, the photons arrival rate can be less than 1 per time constant, and with a suitable readout system, photon counting can provide a means of overcoming electronic noise and the low-frequency TLS device noise.  In this mode, the performance metric is not noise equivalent power but dark counts or extraneous signals that are not coincident across many detectors simultaneously (coincidences can be identified and removed).  Examples might include scattered optical photons, secondary low energy events from cosmic rays, or background radioactivity.  These are typically present larger excitation than the photon pulses, so can be discriminated from them and removed.}

%\textcolor{red}{suggest remove: The intrinsic energy resolution of the detector offers the possibility of discriminating against such signals, and %extraneous signals, for example, scattered optical photons or secondary low energy events from cosmic rays or background radioactivity that may not be observable across many detectors simultaneously.  
%and good energy resolution will be important for achieving a low background count rate.  The ultimate limiting resolution of a pair-breaking detector is set by the Fano limit:
%\begin{equation}
%    R_\mathrm{FANO} = \frac{1}{2\sqrt{2\log{2}}} \sqrt{ \frac{\eta_{pb} E_\gamma}{\Delta F}} \approx 8,
%\end{equation}
%with Fano factor $F \approx 0.2$.  The reduced experimental $R$, along with the reduced measured $\eta_{pb}$ suggest that energetic phonon loss\cite{kozorezov2007electron} is a contributing element to the response of the detector, which is not surprising as the photon energy is similar to the Debye energy, so the phonon-controlled phase of the energy cascade process is similar to that which occurs in detectors for visible and higher energy photons where energetic phonon loss clearly occurs\cite{de2021phonon, zobrist2022membraneless, kozorezov2008phonon}.}

At higher optical power, corresponding to bright sources, the detector measures the integrated photon energy.  In this regime, stability is a key performance goal for many applications. For example, a possible avenue toward detecting exoplanetary life is through mid-IR transit spectroscopy \cite{staguhn2019ultra,mandell2022mirecle}.  At $\lambda = 25\mu$m, the planet to star contrast ratio for an earth-like planet orbiting an M dwarf star is a few thousand.  The hour-scale stability demonstrated in Fig.~\ref{fig:allan} and Fig.~\ref{fig:mod100K} would allow relevant spectral lines to be measured over the secondary eclipse.  As shown in Appendix \ref{sec:res}, the detector can easily modified to work in the 5 to 20 micron range most relevant for mid-IR transit spectroscopy. 

The ultimate limit for KIDs in total power mode at low modulation frequencies is two-level systems (TLS) \cite{gao2007noise} in the capacative elements of the resonator, which create device-level frequency fluctuations indistinguishable from optical power.  %from  which cause the resonant freuquency to fluctuate.  KIDs are based on a measurement of frequency, a physical quantity that may be measured with very high fidelity.  
%The frequency of superconducting resonators is known to fluctuate due to two-level systems (TLS) in the capacitive elements of the resonator \cite{gao2007noise}.  
By measuring the readout power dependence of the quality factor at low readout powers, we determined the intrinsic TLS loss factor $F \delta_0 \approx 10^{-5}$ which points to a significant TLS population, consistent with our measured noise.   However, the effect of TLS noise is
\textcolor{black}{mitigated in our devices by the} high photon response. \textcolor{black}{Additional improvements are possible, as our measured frequency fluctuation noise is actually poorer than observed with larger, less responsive KIDs.  A promising avenue is to simply change the capacitor metal to one with better surface quality.  Vissers et al \cite{Vissers_20} show frequency fluctuation variance a factor of 100 lower than our measured value at 1 Hz using titanium nitride capacitors.  If paired with our inductor, this improved TLS noise level would produce a factor of 10 further improved sensitivity at low modulation frequencies.}  Another avenue to reduce the TLS noise is to increase the responsivity further by decreasing the absorber volume at the cost of reduced dynamic range.  Calculations of the focused optical spot size suggest that the absorber diameter may be reduced by a factor of 2 with only a few percent efficiency loss, improving the responsivity by a factor of 4.
%By changing the capacitor metal to one with better surface quality, it is likely that the TLS effects might be reduced and the detector stability further improved.

Our measurements and understanding of the black body setup suggest a low coupling efficiency $\eta_o = 0.46$.  Based on our validation of the absorber design described in Section~\ref{sec:res} and simulations using the measured lenslet profile, this issue does not seem to be fundamental.  Optimization of the lenslets and minimization of the thickness of the epoxy bond layer between detector and lenslet array substrates can be expected to yield $\eta_o \gtrsim 0.9$.  The addition of a back short located $\lambda/4$ behind the absorber would also improve the detector response, increasing the peak $\mathcal{A}_{abs}(\nu)$ from $\sim 0.7$ to close to 1.

\section{Conclusion}\label{sec:conclusion}

The detector design presented above extends the state-of-the-art for mid- to far-IR photon detection by providing a solution for detector arrays with thousands to tens of thousands of pixels that simultaneously meet \textcolor{black}{the three cardinal performance requirements of  future instruments and experiments.  Our devices show: a) zodi-limited sensitivity for moderate-resolution (R$\sim$100) spectroscopy, including photon counting at low powers, b) natural dynamic range suitable for virtually any envisioned astronomical measurement, c) stability to offer shot-noise limited performance even on modulation timescales of hours.  Further extensions are possible.  With lower-noise readout amplifiers, the single-photon capability of these detectors should be extensible out to much longer wavelengths, enabling for example photon counting with very-high-spectral-resolution (e.g.\ R$\sim10^5$) instruments in the 100--200 micron range.  The low frequency stability can be further improved with demonstrated approaches to reduce TLS noise. }

\backmatter

\bmhead{Supplementary information}

This article has accompanying supplementary information. 

\bmhead{Acknowledgements}

The research was carried out at the Jet Propulsion Laboratory, California Institute of Technology, under a contract with the National Aeronautics and Space Administration (80NM0018D0004).  We acknowledge support from the NASA Strategic Astrophysics Technology (SAT) program under grant number 141108.04.02.01.70 to C.M. Bradford et al.  The authors thank the NASA GSFC microlens team, Manuel Quijada, Jessica Patel, Ed Wollack and Johannes Staguhn. The microlens fabrication and hybridization and the FTS measurements were supported in part by internal research and development awards at NASA GSFC. NFC was supported by an NASA Postdoctoral Program Fellowship at NASA GSFC, administered by ORAU.

% \putbib
% \end{bibunit}

% \newpage
% \begin{bibunit}

% \begin{center}
% \Large{Supplementary Information for:  ``A 25-micron Single Photon Sensitive Kinetic Inductance Detector''}\\
% \end{center}

\begin{appendices}

\section{$\delta x / \delta N_{qp}$}
\label{sec:dxdNqp}

The response of the fractional frequency $\delta x$ to changes in the quasiparticle density can be written as 
\begin{equation}
    \frac{\delta x}{\delta N_{qp}} = \frac{1}{2}\alpha\gamma_s S_2(f_{res}, T) \frac{1}{V n_c} ,
\end{equation}
where $\alpha$ is the ratio of the kinetic inductance to the total inductance of the resonator.  Based on Sonnet simulations of the inductor geometry and comparison with the experimentally measured resonant frequencies, we find $\alpha = 0.81$.  The parameter $\gamma$ characterizes the scaling of the surface impedance with the complex conductivity $Z_s \sim \sigma^{-\gamma}$.  For our 40~nm thick aluminum film, we estimate a mean free path $l \approx 27$~nm based on a measured sheet resistance $R_\square = 0.38\Omega$.  The effective penetration depth is then $\lambda \approx \lambda_L \sqrt{\xi_0 / l} = 119\,\mathrm{nm}$, so the thin film limit applies and $\gamma \approx 1$. 

The function $S_2(f_{res}, T)$ defined in ref \cite{zmuidzinas2012superconducting} contains the Mattis-Bardeen equation for $\sigma_2$.  The value of the function is weakly frequency and power dependent.  For $f_{res} = 600$~MHz and $T = 0.15$~K, $S_2 = 3.96$.  The parameter $n_c = 2 N_0 \Delta$ can be thought of as the density of Cooper pairs.  

Using $N_0 = 1.72\times 10^4 / \,\mu\mathrm{eV}\mu\mathrm{m}^3$ for the single spin density of states at the Fermi level and the value of the gap parameter $\Delta = 0.21$~meV determined by fitting the resonator frequency temperature dependence, we find $n_c = 7.3\times 10^6 / \mu\mathrm{m}^3$.  Using $V = 11\, \mu\mathrm{m}^3$, we arrive at $\delta x / \delta N{qp} = 2\times 10^{-8}$.

\section{Resonant photon absorber}\label{sec:res}

Techniques for absorbing optical energy in the sensitive inductive element of a KID include the use of antenna structures \cite{day2006antenna, baryshev2011progress} or direct absorption in an area-filling metal structures \cite{doyle2008lumped}.  

For the direct method, the absorber takes the form of a meandering trace and doubles as the inductive element of the resonator.  At Far-IR wavelengths, the effective surface impedance of the meander should be adjusted to satisfy an impedance matching condition.  For example, in the case of illumination through a silicon substrate and with a conductive back short located at a distance $\lambda/4$ behind the absorbing surface, the optimal surface impedance is $Z_s = 377/\sqrt{\epsilon_{\mathrm{Si}}}\,\Omega$.  Without a back short some power is lost to transmission and the optimal efficiency is \textcolor{black}{77\%} with \textcolor{black}{$Z_s = 377/(1+\sqrt{\epsilon_{\mathrm{Si}}})\,\Omega$}.

A parallel array of closely spaced wire traces with sheet resistance $R_s$, period $p$ and width $w$ approximates a uniform conducting surface for the polarization with E-field parallel to the traces.  For $\rho = p\sqrt{\epsilon_{\mathrm{Si}}}/\lambda \ll 1$,
$Z_s \approx pR_s/w$ is real, but $Z_s$ begins to develop a significant imaginary part at $\rho \sim 0.3$ and at $\rho = 1$ power is scattered into grating modes.  At short wavelengths, the requirement on $\rho$ implies the need for either a resistive film on the order of $R_s \sim 10$s of ohms or very narrow traces.  For a silicon substrate and $\lambda = 25$~\micron, requiring $\rho < 0.3$ implies a wire period of $p<2.2$~\micron.  An efficient absorber could be made using a high resistivity material like TiN \cite{perido2023parallel}.  For example a 30~nm thick film with $R_s = 50$~Ohms would need an easily fabricated $w\sim 1$~\micron.  The resulting absorber volume would be quite large and only suitable for relatively high-background applications.

Use of a more conductive metal film, such as aluminum, results in a low volume by forcing the traces to be much narrower.  But for a typical aluminum film sheet resistance of $R_s \sim 1$~Ohm, the needed $w \approx 20$~nm, which would be very challenging to fabricate and may shift the detector operation into a ``geiger'' or SNSPD mode because of the small superconducting condensation energy per square of such a trace.  

Another complication with conductive metals is that the surface impedance becomes complex at high frequency where the normal state skin depth becomes comparable to or shorter than the film thickness.  For our $t = 40$~nm thick aluminum film with cold resistivity $1.5~\mu\Omega\cdot$cm, the skin depth $\delta$ is 18~nm at 12~THz, and 
% \begin{equation}
%     Z_s = \frac{k}{\sigma}\coth{kt} = (0.82 + 0.86i)\Omega,
%     \label{eqn:Zs}
% \end{equation}
% where $\sigma$ is the conductivity, $k = (1 + i)/\delta$.  
the imaginary part of $Z_s$ is comparable to the real part.  Because of the imaginary part, an impedance match would only be achieved with the use of a closely spaced (capacitive) backshort.

Our solution to impedance matching to aluminum is to make use of resonant structures, similar to what is done in circuit design for matching between unequal or complex impedances.  The repeated ``hairpin'' structure shown in panel D of Fig.~\ref{fig:pixel} has a resonant response to waves polarized both parallel and perpendicular to the direction of the hairpin, and the details of the geometry can be tuned to adjust the resonance frequencies for the polarizations independently.  For example the length of the hairpin affects mainly the parallel polarization, while the location of the connector that is offset from the end of the hairpin affects mainly the perpendicular polarization resonance.

\begin{figure}[h]
\centering
\includegraphics[width=1.00\textwidth]{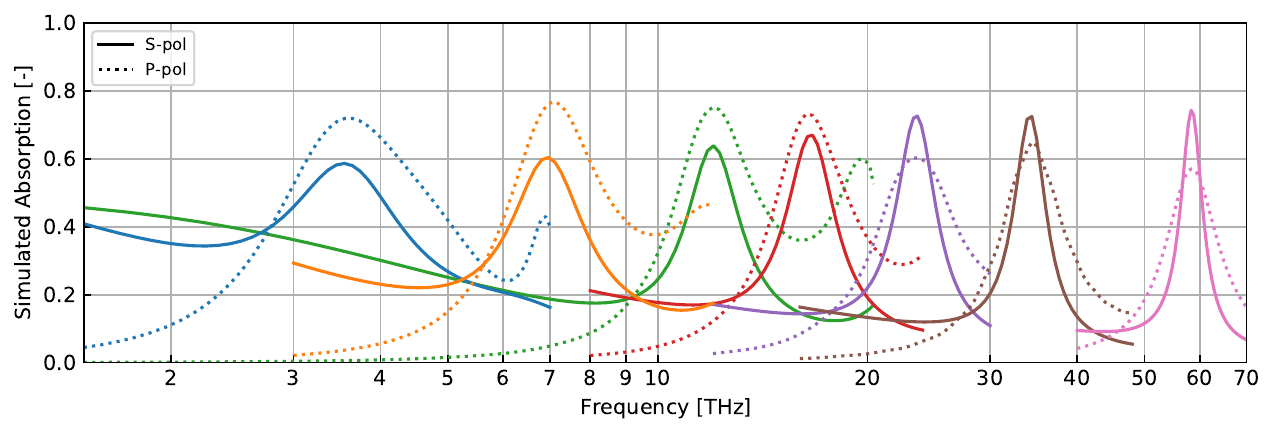}
\caption{Simulated absorption efficiency for absorbers designed for 3.5, 7, 12, 24, 34 and 58~THz.  The P-polarizaton is in the direction parallel to the hairpin, while the S-polarization is along the connected path of the meander structure. The line width of the hairpin structure is scaled with frequency from 200~nm at 3.5~THz to 60~nm at 58~THz.}
\label{fig:absorbers}
\end{figure}

%We simulate the absorption efficiency of the structures in HFSS considering an infinite array of the repeat structures and using Equation~\ref{eqn:Zs} for the surface impedance.  Results for a range of geometries are shown in Fig.~\ref{fig:absorbers}.  The simulations did not include back shorts, so the maximum expected efficiency is 77\%, close to the simulated results.  Including a back short, the peak absorption can be increased to greater than 90\%.  These simulations suggest that this absorber design can be applied over a wide frequency range over the Far- and Mid-IR bands.  Silicon has some absorption bands in the Mid-IR which could be avoided by switching to a germanium substrate.

The absorption efficiency of the structures was simulated in HFSS by considering an infinite array of the repeat structures.  The surface impedance of a finite thickness film with conductivity $\sigma$ is given by Kerr\cite{KerrMemo245},
\begin{equation}
    Z_s = \frac{k}{\sigma}\frac{ \exp(kt) + \frac{\sigma Z_\eta - k}{\sigma Z_\eta + k} \exp(-kt) }{ \exp(kt) - \frac{\sigma Z_\eta - k}{\sigma Z_\eta + k} \exp(-kt) },    
\end{equation}
where $t$ is the film thickness, $k = (i\omega\sigma\mu_0)^{1/2}$ and $Z_\eta = 377\Omega$ is the terminating impedance on the side of the film opposite the incident wave.  When the frequency becomes comparable to the rate of electron scattering events, the conductivity is modified according the Drude formula,
\begin{equation}
    \sigma(\omega) = \frac{\sigma(0) }{1 + i\omega\tau},
\end{equation}
where the scattering time $\tau = 13$~fS for our aluminum film.
As suggested by Kerr\cite{KerrMemo245}, the finite thickness metal film can be modeled as two sheets separated by $t$ with effective surface impedance 
\begin{equation}
    Z_x = \frac{1}{2}\left[ (2Z_s - i\omega\mu_0t) + \sqrt{ 4Z_s^2 + (i\omega\mu_0t)^2 } \right].    
\end{equation}

Results for a range of geometries are shown in Fig.~\ref{fig:absorbers}.  The simulations did not include back shorts, so the maximum expected efficiency is 77\%, close to the simulated results.  Including a back short, the peak absorption can be increased to greater than 90\%.  These simulations suggest that this absorber design can be applied over a wide frequency range over the Far- and Mid-IR bands.  Silicon has some absorption bands in the Mid-IR which could be avoided by switching to a germanium substrate.

\begin{figure}[h]
\centering
\includegraphics[width=1.00\textwidth]{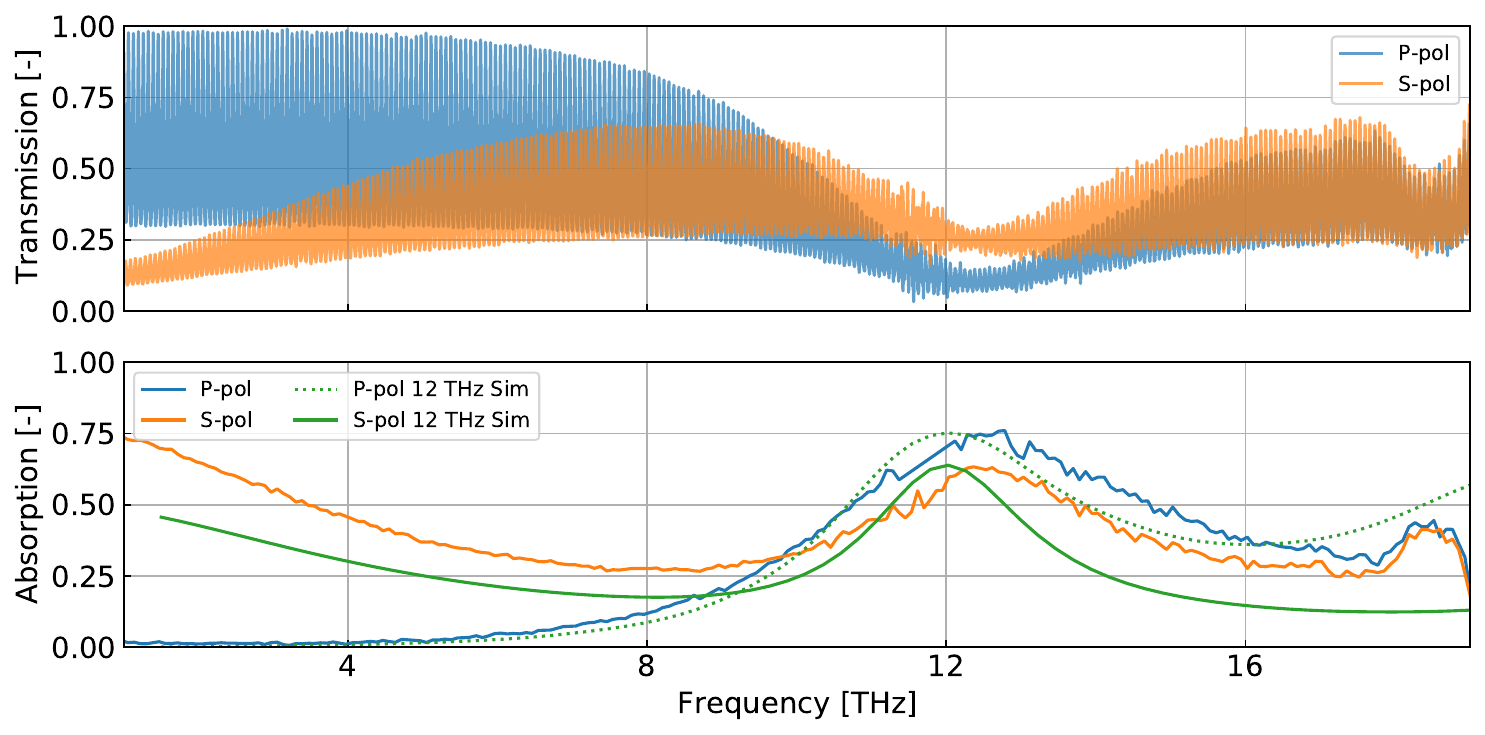}
\caption{{\it Top:}. Polarized transmission measurement of the absorber test sample.  S-pol is parallel to the hairpin and P-pol is orthogonal.  {\it Bottom:}  Extracted absorption compared to the simulation result.}
\label{fig:Absorber_FTS}
\end{figure}

In order to experimentally validate the spectral efficiency of the absorber pattern used for the detectors described in the paper, samples were fabricated at JPL with the pattern continuously repeated over a 2~cm by 2~cm area.  These samples were cooled to 5~K and measured in transmission in an FTS at NASA Goddard.  The raw transmission shown in the upper panel of Fig.~\ref{fig:Absorber_FTS} shows Fabry-Perot fringes caused by the cavity of the silicon substrate. The peaks and nulls of the asymmetric Fabry-Perot interferometer (FPI) can be used to determine the absorptance of the patterned mesh \cite{perido_kinetic_2023}. The transmission of an asymmetric FPI (consisting of a silicon cavity with bare silicon on one side and an absorber mesh on the other) can be written as
\begin{equation}
    T = \left|\frac{E_t}{E_0}\right|^2 = \left|\frac{t_{Si}t_{abs}\exp(i\beta L)}{1-r_{Si}r_{abs}\exp(2i\beta L)}\right|^2,
    \label{eq:asymm_fpi}
\end{equation}
where $L$ is the silicon substrate thickness, $\beta=2\pi\nu/n_{Si}c$, and  $t_{Si}=\frac{2}{1+n_{Si}}$ and $r_{Si}=\frac{n_{Si}-1}{n_{Si}+1}$ are the amplitude transmission and reflection coefficients at the silicon surface of the cavity. The mesh absorptance is related to the mesh transmission and reflection coefficients by
\begin{equation}
    \mathcal{A}_{abs}(\nu) = 1 - r_{abs}(\nu)^2 - t_{abs}(\nu)^2/n_{Si} ,
    \label{eq:mesh_abs}
\end{equation}
and we can determine $t_{abs}$ and $r_{abs}$ by evaluating Equation~\ref{eq:asymm_fpi} at the peaks and nulls of the FPI fringes. The fringe maxima and minima occur when $\exp(i\beta L) = +1$ and $-1$, respectively, such that
\begin{equation}
    T_{max} = \left|\frac{t_{Si}t_{abs}}{1-r_{Si}r_{abs}}\right|^2 ~~~~~{\rm and}~~~~~ T_{min} = \left|\frac{t_{Si}t_{abs}}{1+r_{Si}r_{abs}}\right|^2.
\end{equation}
Solving the system of equations for $t_{abs}$ and $r_{abs}$ yields:
\begin{equation}
    t_{abs} = \frac{2}{t_{Si}}\left(\frac{1}{\sqrt{T_{max}}}+\frac{1}{\sqrt{T_{min}}}\right)^{-1} ~~~{\rm and}~~~ 
    r_{abs} = \frac{1}{r_{Si}}\left(\frac{1-\sqrt{T_{max}/T_{min}}}{1+\sqrt{T_{max}/T_{min}}}\right).
    \label{eq:mesh_coeffs}
\end{equation}

By using the peaks and nulls of the data in the upper panel of Fig.~\ref{fig:Absorber_FTS} for $T_{max}$ and $T_{min}$ and Equation~\ref{eq:mesh_coeffs}, we determine the mesh absorption using Equation~\ref{eq:mesh_abs}, as plotted in the lower panel of Fig.~\ref{fig:Absorber_FTS}. The simulated mesh absorption is overplotted on top of the data. We find that the center of the resonance is shifted to slightly higher frequency than designed (12 THz), but that the peak absorption and resonance bandwidth are roughly as expected.

\section{Base temperature response} \label{sec:basetresp}

 A measurement of the resonator frequency versus base temperature is used to determine superconducting gap parameter $\Delta_0$ by fitting to the Mattis-Bardeen expression for the temperature dependence of the complex conductivity.  To avoid fitting simultaneously to both $\Delta_0$ and the kinetic inductance fraction $\alpha$ of the resonator, we first determine $\alpha$ by comparing the measured resonance frequencies to Sonnet simulations of the resonator, varying the aluminum absorber sheet inductance until the simulated resonance frequency matches.  Using this procedure, we determine $\alpha = 0.81$.  

\begin{figure}[h]
\centering
\includegraphics[width=0.52\textwidth]{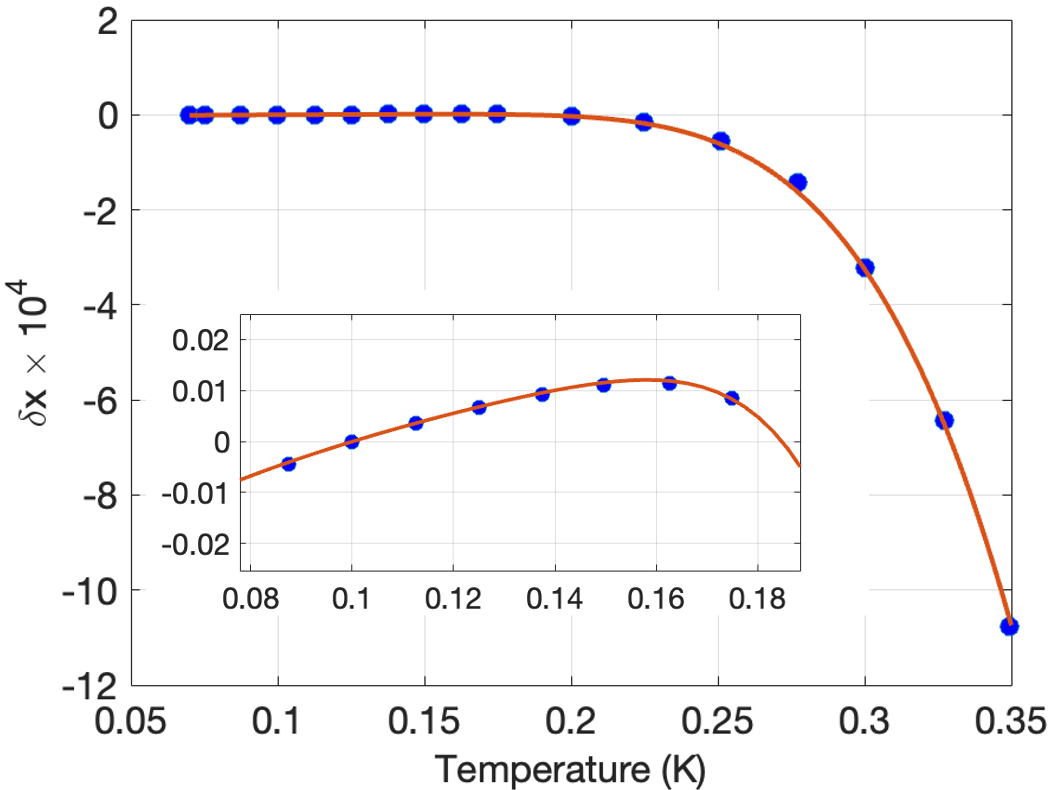}
\includegraphics[width=0.44\textwidth]{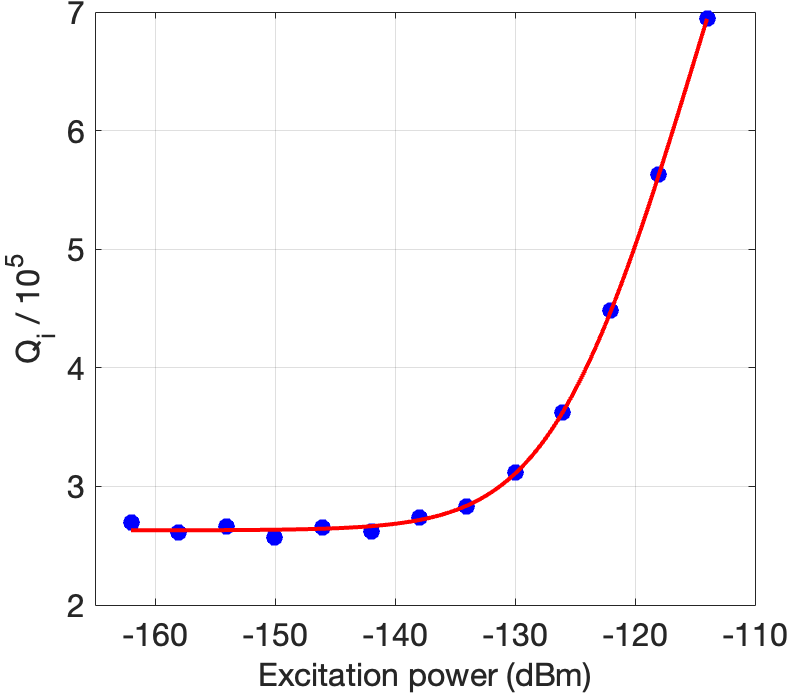}
\caption{Left: Fractional frequency shift of the 602~MHz resonator versus base temperature.  The red line is a fit to the Mattis-Bardeen temperature dependence including a TLS contribution.  {\it Inset:}  Detail of the low temperature behavior. Right: Measured internal quality factors $Q_i$, derived from fits to VNA scans of the 602~MHz resonance, as a function of drive power.}
\label{fig:dxvt_and_QivP}
\end{figure}

% \begin{figure}[h]
% \centering
% \includegraphics[width=0.7\textwidth]{figures/dxvtcomp.png}
% \caption{Fractional frequency shift of the 602~MHz resonator versus base temperature.  The red line is a fit to the Mattis-Bardeen temperature dependence including a TLS contribution.  {\it Inset:}  Detail of the low temperature behavior.}
% \label{fig:dxvt}
% \end{figure}

%  \begin{figure}[h]
% \centering
% \includegraphics[width=0.6\textwidth]{figures/QivP.png}
% \caption{Measured internal quality factors $Q_i$, derived from fits to VNA scans of the 602~MHz resonance, as a function of drive power.}
% \label{fig:QivP}
% \end{figure}

The fractional frequency shift with temperature is shown in the left panel of Fig.~\ref{fig:dxvt_and_QivP}.  At low temperatures, the resonance shifts to higher frequency with increasing temperature due to the influence of two-level-systems (TLS) at interfaces \cite{anderson2012amorphous}.  Both the dielectric temperature dependence and the low-temperature, low-power dielectric loss are connected to the product of the TLS filling factor and the intrinsic TLS loss tangent $F\delta_0$, which we determine by measuring the resonator internal quality factor $Q_i$ versus excitation power (Right panel of Fig.~\ref{fig:dxvt_and_QivP}).

The measured $Q_i$ versus power are fit to the TLS loss power dependence \cite{hunklinger1976physical}
\begin{equation}
    Q_{i, TLS}^{-1} = F\delta_0 \frac{\tanh(hf_{res}/2k_b T)}{\sqrt{1 + P/P_c}} + Q_x^{-1},
\label{eqn:Qitls}
\end{equation}
where $f_{res}$ is the resonator frequency, $Q_x$ and additional loss mechanism of undetermined origin, and $P_c$ can be related to a characteristic electric field $E_c$ of the TLS material.  The measurement was done at 70~mK, which was the lowest temperature accessible on the stage of the cryostat on which the sample was mounted.  The fit yields $1/F\delta_0 = 1.0\times10^5$.  The resonator $Q_i$ values are higher than that value due to both the temperature dependence and the power dependence in Equation~\ref{eqn:Qitls}.  That value was then used in the model for the TLS contribution to the temperature dependent frequency shift \cite{hunklinger1976physical}
\begin{equation}
    \frac{\delta\epsilon}{\epsilon} = \frac{2F\delta_0}{\pi} \left\{ \Re \left[ \Psi \left( \frac{1}{2} - \frac{hf_{res}}{2\pi i k_b T} \right)  \right] - \log \left( \frac{hf_{res}}{k_bT} \right) \right\},
\end{equation}
with $\Psi$ is the complex digamma function, which matches the data reasonably well over an interval of the low temperature range.  The $\delta x(T)$ data are then fit to a single parameter $\Delta$ yielding 0.21~meV.

\begin{figure}[h]
\centering
\includegraphics[width=0.51\textwidth]{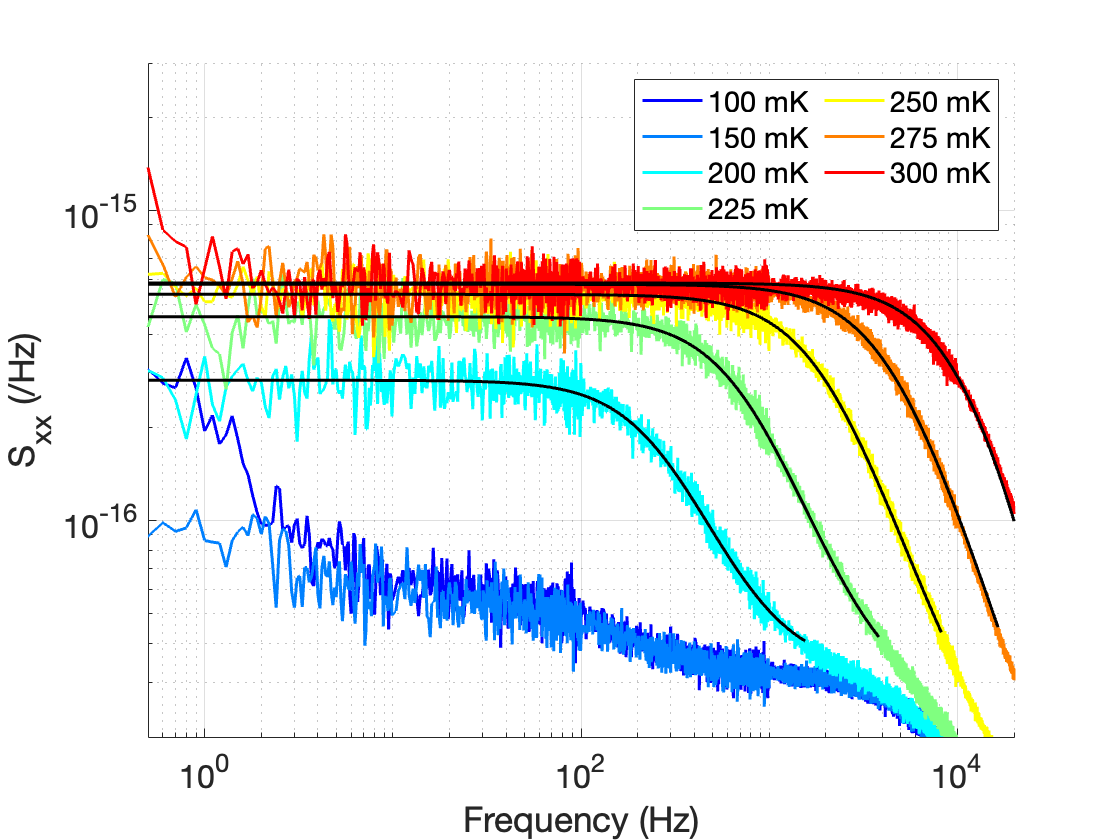}
\includegraphics[width=0.48\textwidth]{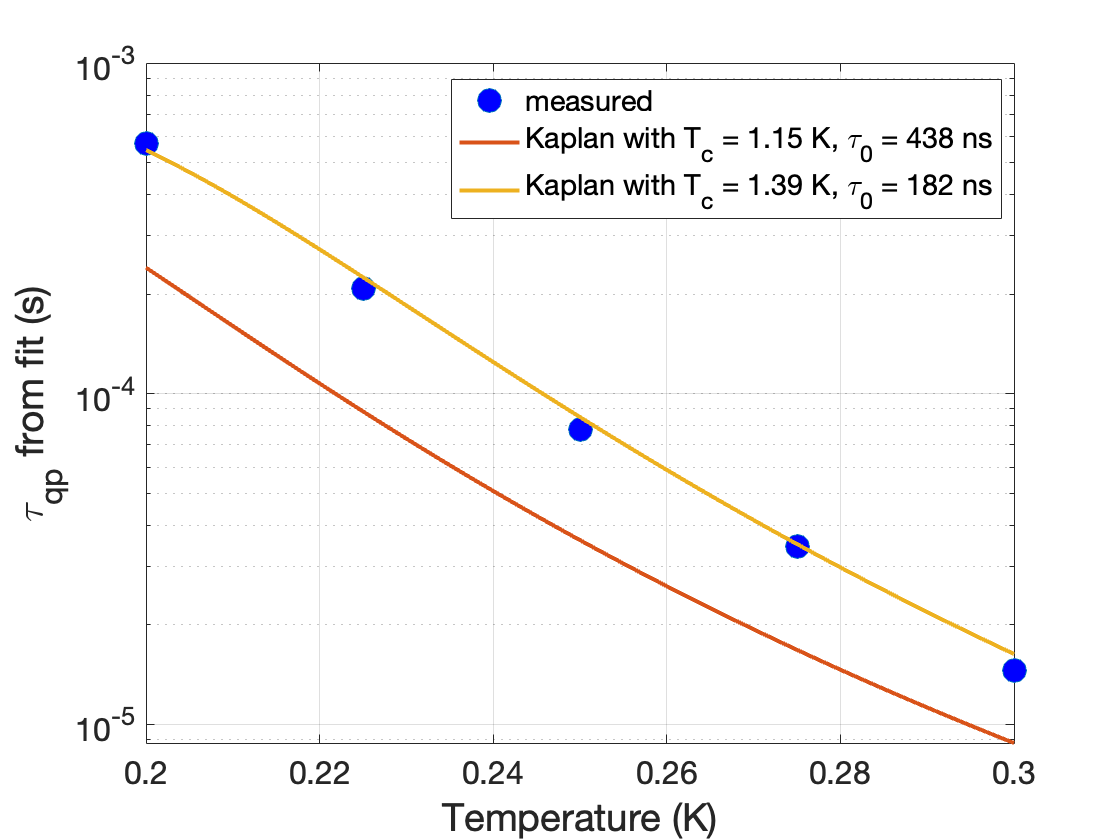}
\caption{Left: Fractional frequency noise measured at several base temperatures.  The solid lines are fits to a single pole roll-off. Right: Quasiparticle lifetimes extracted from noise fits.}
\label{fig:Sxx_and_tau_vs_Tbath}
\end{figure}

The fractional frequency noise for several base temperatures is shown in 
the left panel of Fig.~\ref{fig:Sxx_and_tau_vs_Tbath}.  The fits to these noise PSDs are used to derive the recombination time of thermally generated quasiparticles shown in the right panel of Fig.~\ref{fig:Sxx_and_tau_vs_Tbath}.  The solid lines in that figure are predictions of the Kaplan theory \cite{kaplan1976quasiparticle} for two choices of parameters.  The red line uses the $T_c$ and gap values for bulk aluminum along with characterisic time $\tau_0 = 438$~ns given by Kaplan.  A better fit is found using the measured gap parameter determined above and the corresponding BCS $T_c$ value along with $\tau_0 = 182$~ns.  The recombination constant 
\begin{equation}
    R = \left( \frac{2\Delta}{k_b T_c} \right)^3 \frac{1}{4 N_0 \Delta \tau_0} = 16.5 \;\frac{\mu\mathrm{m}^3}{\mathrm{s}},
\end{equation}
which is close to the value found in Section~\ref{sec:etaqpnoise}.

%\ifcomment\textcolor{red}{Plots showing low frequency noise at lower temperatures, which has a bearing on choosing operating temperature. }\fi

\section{Uniformity across the array}\label{sec:uniformity}

\begin{figure}[h]
\centering
\includegraphics[width=0.99\textwidth]{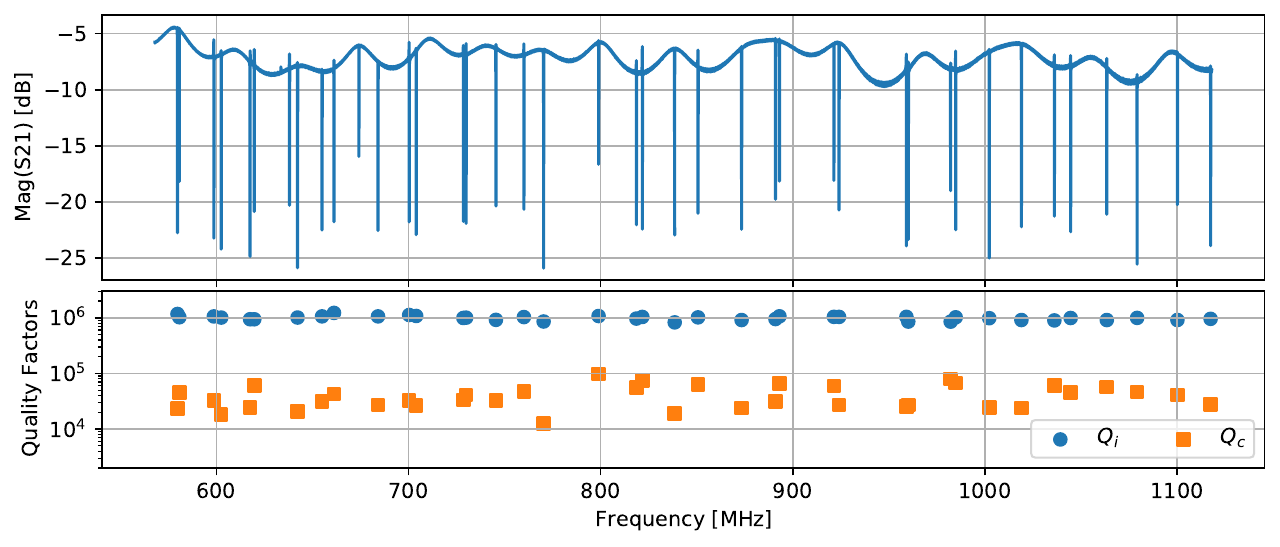}
\caption{Upper panel: S21(f) measurement of the 44 pixel array shows a yield of 41 resonators between roughly 570 and 1120 MHz. Lower panel: Distribution of internal and coupling quality factors ($Q_i$ and $Q_c$, respectively) as a function of resonator frequency.}
\label{fig:VNA_and_Qs}
\end{figure}

The upper panel of Fig.~\ref{fig:VNA_and_Qs} shows the S21(f) transmission through the 44-element KID array. The achieved resonator yield was 41/44 (93\%). Each resonator was fit to a non-linear resonator model \cite{swenson2013operation,dai2022new}. The extracted internal and coupling quality factors ($Q_i$ and $Q_c$, respectively) are shown in the lower panel of Fig.~\ref{fig:VNA_and_Qs}. The quality factors are all within their designed ranges. The resonator that was studied in depth in the results presented in this manuscript was located at 602 MHz and had very similar properties to the rest of the resonators in the array.

\begin{figure}[h]
\centering
\includegraphics[width=0.99\textwidth]{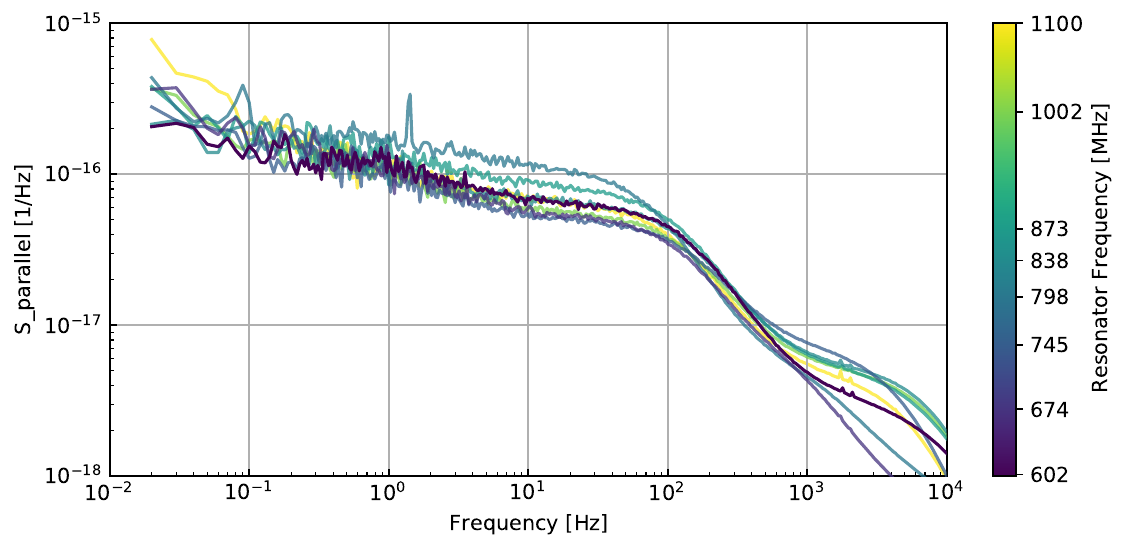}
\caption{Fractional frequency noise spectra of eight optical resonators. The incident power on the microlens of each detector was 0.2 aW.}
\label{fig:Sxx_fs30kHz_allres}
\end{figure}

Fig.~\ref{fig:Sxx_fs30kHz_allres} shows the factional frequency noise power spectra for eight different resonators in the array, measured with incident optical power of 0.2 aW. The 602 MHz that was studied in depth in the results presented above is shown in dark purple. The spectra of all eight resonators show similar properties including white noise level, quasiparticle recombination roll-off, and low-frequency TLS noise ($S_{TLS}\propto f^{1/4}$).

The black body coupling was found to be similar across these same eight resonators. The mean photon count rate with the black body at 30~K was 5.15 s$^{-1}$ with a standard deviation of 0.28 s$^{-1}$, implying the efficiency factor $\eta_o$ shown in Fig.~\ref{fig:ratecal} is uniform across the device.

\begin{figure}[h]
    \centering
    \includegraphics[width = 0.99\textwidth]{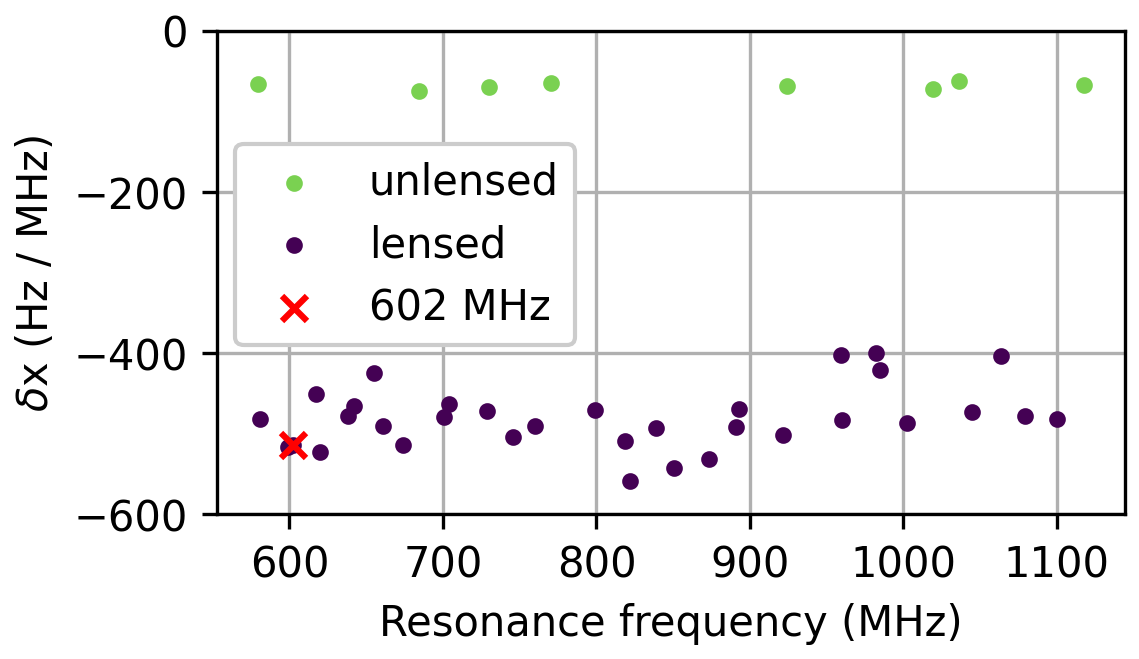}
    \caption{Fractional frequency shift versus resonance frequency for 41 KIDs. The incident power on each detector was 220 fW. The 602 MHz KID under study in this paper is plotted as a red X.}
    \label{fig:resp_uniformity}
\end{figure}

\par To further assess uniformity of the responsivity, fractional frequency shifts were determined for all 41 resonators between two blackbody powers (53 zW and 220 fW). The results are plotted in Fig.~\ref{fig:resp_uniformity}. The 8 KIDs without microlenses are clearly separated from the 33 KIDs with microlenses. The lensed KIDs have an average fractional frequency shift of $7\times$ the unlensed KIDs, which corresponds to a detected power ratio of $\sim 50$.  That ratio is smaller than the ratio of the lenslet are to the bare absorber area $(870/70)^2 = 150$. The observed photon rate on the unlensed 684 MHz resonator was measured at 40~K, 42~K, 44~K, and 45~K, with corresponding efficiencies for the predicted photon flux $\Gamma_{\gamma}$ (Eq. \ref{eq:photon_rate}) found to be $\eta_o$ = 1.04, 1.00, 1.00, and 0.92, respectively. This suggests that the microlens is acting nonideally, as the predicted flux model describes loading on the bare detectors with $\eta_o$ near unity. However, the decreasing trend in the dark efficiencies may indicate contributors subdominant to the microlens, such as an aperture thickness effect, scattering from the other metal structures of the pixel, or other stray light reflected within the silicon substrate.

\section{Coincidence Triggering}\label{sec:ctrigger}

Two homodyne circuits were combined to collect simultaneous $S_{21}$ measurements of the 602 MHz and 1079 MHz resonators. These resonators are spatially directly adjacent, with a spacing of 900~\micron\ between their absorbers. A nearest neighbor pair was chosen to maximize the likelihood of a non-photonic event, such as a cosmic ray, exciting both resonators. A 25~\micron\ photon will only be seen as a pulse in a single resonator.

\begin{figure}[h]
    \centering
    \includegraphics[width = 0.67\textwidth]{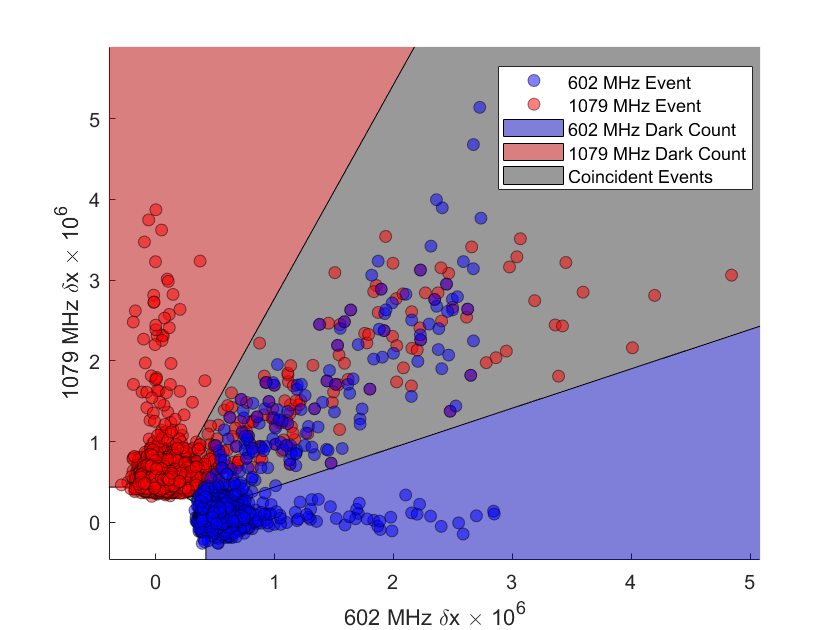}
    \caption{Simultaneous measurements of the factional frequency shift for the 602 MHz and 1079 MHz resonators over 8000~seconds of data.}
    \label{fig:dcr}
\end{figure}

Fig. \ref{fig:dcr} shows the amplitudes of pulse-like events detected in either resonator. Blue points indicate events seen in the 602 MHz resonator, with corresponding $\delta x$ value for the 1079 MHz resonator taken to be the maximum value in a 40~$\mu$s window around the time the peak of the pulse seen in the 602 MHz resonator. Red points are similarly defined for the events seen in the 1079 MHz resonator. The point distribution can be divided into a low-amplitude noise ball, a central region of shared pulses belonging to cosmic ray events, a horizontal lobe for photon pulses in the 602 MHz resonator, and a vertical lobe for photon pulses in the 1079 MHz resonator. The shared event region is defined by extending 3 standard deviations above and below the best fit line to the central data outside the noise ball. Setting a cutoff amplitude and counting pulses in the horizontal lobe gives the DCR in table~\ref{tab:dcrtable}.

\end{appendices}


\bibliography{sn-bibliography}% common bib file
% \putbib
% \end{bibunit}

\end{document}